\documentclass[twocolumn,twocolappendix]{aastex7}

\usepackage{amsmath} 
\usepackage[style=iso]{datetime2} 
\usepackage{booktabs} 
\usepackage{ragged2e} 
\usepackage{needspace} 
\usepackage{longtable} 
\usepackage{xcolor} 
\definecolor{plum}{HTML}{9e0d66}
\definecolor{black}{HTML}{000000}

\newcommand{\planck}{{\it Planck}}
\newcommand{\hersc}{{\it Herschel}}
\newcommand{\HI}{H{\sc i}}

\newcommand{\CII}{[C{\sc ii}]}
\newcommand{\Cplus}{C$^+$}

\makeatletter 
\newcommand\footnoteref[1]{\protected@xdef\@thefnmark{\ref{#1}}\@footnotemark}
\makeatother

\renewcommand{\baselinestretch}{0.925} 

\submitjournal{and published in the Astronomical Journal: \url{https://iopscience.iop.org/article/10.3847/1538-3881/add40f}}

\begin{document}

\title{Measuring Interstellar Carbon Abundance via 158\,\micron\ \CII\ Absorption with SOFIA \linebreak-- \textcolor{black}{A Potential} Detection, and Proof-of-Concept for Depletion Studies with Future Far-IR Facilities}

\correspondingauthor{Christopher J. R. Clark}
\email{cclark@stsci.edu}


\author[0000-0001-7959-4902]{Christopher J. R. Clark}
\affiliation{AURA for the European Space Agency, Space Telescope Science Institute, 3700 San Martin Drive, Baltimore, Maryland 21218, USA}
\email{cclark@stsci.edu}

\author[0000-0001-6326-7069]{Julia. C. Roman-Duval}
\affiliation{Space Telescope Science Institute, 3700 San Martin Drive, Baltimore, Maryland 21218-2463, USA}
\email{duval@stsci.edu}

\author[0000-0003-3229-2899]{Suzanne C. Madden}
\affiliation{AIM, CEA, CNRS, Université Paris-Saclay, Université Paris Diderot, Sorbonne Paris Cit\'e, 91191 Gif-sur-Yvette, France}
\email{suzanne.madden@cea.fr}

\author[0000-0002-4903-9542]{Marc Mertens} 
\affiliation{Max-Planck-Institut für Radioastronomie, Auf dem Hügel 69, D-53121 Bonn, Germany}
\affiliation{I. Physikalisches Institut, Universität zu Köln, Zülpicher Straße 77, 50937 Köln, Germany}
\email{mertens@ph1.uni-koeln.de}

\author[0000-0002-7743-8129]{Claire E. Murray} 
\affiliation{Space Telescope Science Institute, 3700 San Martin Drive, Baltimore, Maryland 21218-2463, USA}
\affiliation{Department of Physics \& Astronomy, Johns Hopkins University, 3400 N. Charles Street, Baltimore, MD 21218, USA}
\email{cmurray1@stsci.edu}

\author[0000-0001-7658-4397]{J\"urgen Stutzki} 
\affiliation{I. Physikalisches Institut, Universität zu Köln, Zülpicher Straße 77, 50937 Köln, Germany}
\email{stutzki@ph1.uni-koeln.de}

\author[0000-0003-1356-1096]{Elizabeth Tarantino}
\affiliation{Space Telescope Science Institute, 3700 San Martin Drive, Baltimore, Maryland 21218-2463, USA}
\email{etarantino@stsci.edu}

\author[0000-0003-0789-9939]{Kirill Tchernyshyov}
\affiliation{Department of Astronomy, University of Washington, Seattle, Washington 98105, USA}
\email{ktcherny@uw.edu}



\begin{abstract}

Carbon plays key roles in the InterStellar Medium (ISM) -- as a constituent of dust, as the carrier of the dominant far-infrared cooling line, and as a component of various important molecules. But despite this, there are very few measurements of the abundance and depletion of carbon in the diffuse ISM. As with other elements, these measurements are traditionally performed in the ultraviolet. But for carbon, such measurements are extremely difficult, and less than 20 have been reported in the literature to date. Here, we present a novel method of measuring the abundance and depletion of carbon in the diffuse ISM: by observing absorption of the 158\,\micron\ \CII\ line in the far-infrared. We present a catalog of 432 candidate sightlines that use bright nearby galaxies as background sources, and predict the \CII\ absorption expected towards each. We conducted a pilot study using SOFIA, targeting sightlines towards the galaxies IC\,342 and Circinus. We report a \textcolor{black}{potential} detection of Galactic \CII\ absorption along the IC\,342 sightline, although it requires disentangling \CII\ emission from IC\,342 itself. The Circinus sightline had an insufficiently stable instrumental baseline to allow a detection. This SOFIA study informs the prospects for \CII\ absorption measurements with future facilities. To that end, we explore the potential for four proposed future FIR telescopes -- PRIMA, FIRSST, SALTUS, and {\it Origins} -- to detect \CII\ absorption. We find that all four facilities would be able to detect \CII\ absorption along a significant number of sightlines. 

\end{abstract}



\begin{linenumbers}

\needspace{3\baselineskip} \section{Introduction} \label{Section:Introduction}

It is easy to argue that carbon is the most important metal in the InterStellar Medium (ISM). In the dust phase, carbonaceous grains make up one of the two main grain varieties (along with oxygen-rich silicates; \citealp{Draine2001A,Jones2013C}). The mid-infrared emission of galaxies' ISM is dominated by aromatic and aliphatic carbonaceous features \citep{Draine2007C,Jones2016A,Sandstrom2012B}. At shorter wavelengths, carbon plays a key role in dust extinction. For instance, the prominent far-ultraviolet rise and 2175\,\AA\ bump extinction features are believed to be due to carbonaceous species \citep{Draine2003A,Gordon2003B} -- with evidence that the species causing the far-ultraviolet extinction rise is closely linked to the formation and/or shielding of molecular hydrogen \citep{VanDePutte2023A}. Meanwhile, in the gas phase, the \CII\ 158\,\micron\ fine structure transition is the primary cooling line for the cold ISM, thereby playing a key role in regulating star-formation \citep{Wolfire2008A,Glover2014A}, and being a commonly-observed line for probing the star-forming ISM \citep{Carilli2013D,Croxall2017B,Tarantino2021A}.

However, the fraction of carbon that is in the gas phase, compared to the fraction that is depleted into dust grains, is almost unconstrained. As with most metals, the gas-phase abundance of carbon in the neutral ISM is typically measured via UltraViolet (UV) absorption features.  But in the case of carbon (which primarily exists as \Cplus\ in the ISM; \citealp{Goldsmith2012C}), the only available absorption features are either the very weak intersystem C{\sc ii} transition at 2325\,\AA, or the extremely strong C{\sc ii} transitions at 1036 and 1334\,\AA\ \citep{Morton2003B}. The weak 2325\,\AA\ feature requires formidable sensitivity and spectral resolution to even be detected; and fitting the damping wings of the saturated multi-component 1036 and 1334\,\AA\ features atop a complex stellar continuum is even more rarely attempted \citep{Sofia2004A,Sofia2011A}. 

As such, only 19 abundances for carbon in the cold ISM have been published (\citealp{Sofia2011A} and references therein) -- 13 from the 2325\,\AA\ weak intersystem transition, and 6 from the 1334\,\AA\ strong transition. Moreover, the abundances derived via the two methods disagree severely. Along sightlines where abundances have been estimated via both techniques, they differ by a factor of 1.7 on average (with the differences between measurements ranging from a factor of 3.0 to a factor of 1.1; \citealp{Sofia2011A}).

A consequence of this lack of certainty is that the fraction of carbon depleted onto dust, and especially how this depletion varies with column density, is almost unknown. This is illustrated starkly by Figure~5 of \citet{Jenkins2009B}, where the depletion of carbon at a range of column densities (expressed via the `depletion strength' proxy) is plotted, with the depletion of several other key dust-constituting elements shown for comparison. In contrast with the other elements, the variation of carbon depletion with column density is effectively unquantified. As hydrogen column increases from $10^{19}\,{\rm cm^{-2}}$ to $10^{21}\,{\rm cm^{-2}}$, carbon depletion might increase by a factor of 2.5, or it might be totally unchanged -- both are compatible with the existing data. Moreover, UV measurements of abundances -- for all elements, not just carbon -- become effectively impossible for sightlines with high column densities, due to interstellar extinction towards the OB stars used as background sources. This prevents us from measuring depletions in the highest-density environments, where complex chemical evolution is no doubt occurring.

Understanding elemental depletions, and how they vary with density, is critical to studies of the ISM. Depletion strongly influences both the dust-to-gas ratio, a key metric for understanding the chemical evolution of galaxies \citep{Remy-Ruyer2014A,Roman-Duval2022A}, and how galaxies absorb and reprocess starlight \citep{Bianchi2018A}. Carbon is especially central to these questions, as not only is it the second most abundant metal in the ISM, it is also notably volatile, meaning variation in its phase has the potential to significantly \textcolor{black}{affect} the dust-to-gas ratio. This, in turn, dictates our view of the `dust budget crisis', wherein the amount of dust in galaxies, especially at high redshift, can appear to outstrip known dust sources \citep{Rowlands2014B}. 

Separate from this, there is also the ongoing `carbon crisis', arising from the fact that essentially all dust models require a greater carbon abundance than what is actually measured from young stellar photospheric abundances \citep{Wang2015A}. A proposed solution to the `carbon crisis' is that photospheric carbon abundances underestimate interstellar carbon abundances, due to uneven incorporation of different elements from the ISM in the star formation process -- further reinforcing the importance of accurately measuring the {\it in situ} abundance of interstellar carbon.

In this paper, we present a previously-overlooked approach for measuring the abundance and depletion of carbon in the neutral ISM: specifically, the use of 158\,\micron\ \CII\ absorption. We present the results of a pilot study conducted with the Stratospheric Observatory For Infrared Astronomy (SOFIA; \citealp{Pasquale2018A}), and the prospects for taking advantage of this method with future Far-InfraRed (FIR) telescopes. The {\it Hubble} Space Telescope, and its UV spectrometers, have a finite lifespan, so we are therefore facing the prospect of losing the ability to measure ISM metal abundances in the UV, until the successful launch of the Habitable Words Observatory \citep{Astro2020}. By providing an alternate way to measure the depletions of key elements, and to do so in high-extinction environments inaccessible in the UV, FIR absorption-line spectroscopy with a future FIR mission would be a highly valuable tool. 

\needspace{3\baselineskip} \section{Method} \label{Section:Method}

In the diffuse ISM, the vast majority of gas-phase carbon resides in \Cplus\ \citep{Wolfire2010A,Goldsmith2012C,Beuther2014C,Pineda2017A}, with the transition to C$^{0}$ occurring in a relatively thin layer, surrounding the denser molecular phase \citep{Bisbas2018B,Clark2019A}. 

Despite being one of the most commonly-observed metal cooling lines in {\it emission}, there have been very few observations of 158\,\micron\ \CII\ in {\it absorption} \citep{Falgarone2010A,Graf2012A,Stutzki2013A,Gerin2015A,Langer2016B,Nesvadba2016B,Guevara2020B,Kabanovic2022A}. Moreover, these observations have all been of absorption of \CII\ emission associated with regions of intense star formation, with emitting and absorbing material in the same galaxy (usually the Milky Way). This leads to very complex line profiles -- \textcolor{black}{even when the absorbing \Cplus\ is not physically associated with the \CII\ emitting material, the two cannot be distinguished in velocity space \citep{Langer2016B}.} As a result, the associated hydrogen column cannot be isolated with confidence. This means that a clean determination of the carbon abundance cannot be made. 

\needspace{3\baselineskip} \subsection{\CII\ Absorption and Excitation} \label{Subsection:Absorption_and_Excitation}

In this work, we present the novel methodology of using FIR-bright background galaxies to detect foreground 158\,\micron\ \CII\ absorption by the cold Galactic ISM. By observing \CII\ absorption against the continuum of bright background sources, it should be possible to make a clean detection of the absorption profile, without the confusion of self-absorption. 

It is important to note that only \Cplus\ atoms in the fine-structure ground state can cause \CII\ absorption. This is significant, as the ground and excited states in the \CII\ transition have a relatively modest energy separation of $T^\ast = 92\,{\rm K}$. We therefore need to consider the fraction of \Cplus\ expected to be in the fine-structure ground state, as described by the excitation temperature, $T_{\it ex}$. The ratio of relative populations in the upper versus lower energy states, $\frac{n_{\it u}}{n_{\it l}}$, is given by the Boltzmann factor equation:

\begin{equation}
\frac{n_{\it u}}{n_{\it l}} = \frac{g_{\it u}}{g_{\it l}}\ e^{\,T^\ast/T_{\it ex}}
\label{Equation:Boltzmann_Ratio}
\end{equation}

\begin{figure}
\centering
\includegraphics[width=0.475\textwidth]{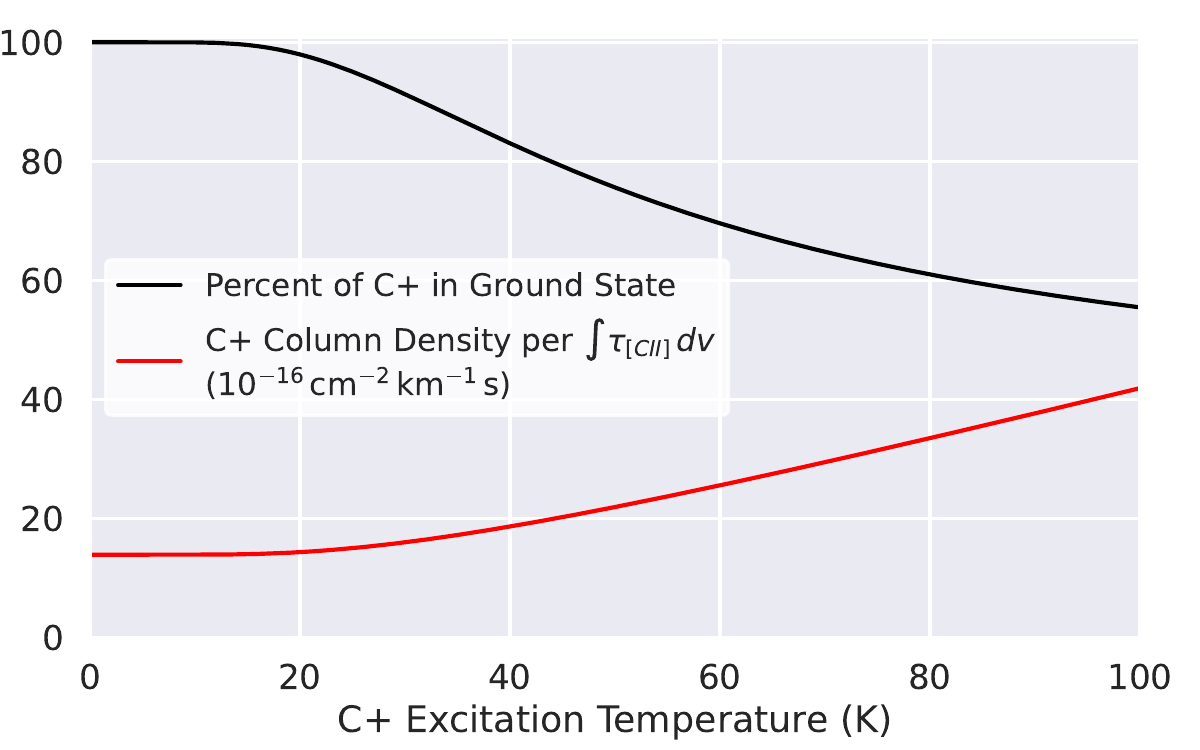}
\caption{\textcolor{black}{{\it Black:} Plot of how the fraction of \Cplus\ ions in the fine-structure ground state evolves with excitation temperature. {\it Red:} Plot of how the column density of \Cplus\ determined from a given level of \CII\ integrated line opacity, $\int \tau_{\rm [CII]}\ dv$, evolves with excitation temperature.}}
\label{Fig:Excitation_Plots}
\end{figure}

\noindent\textcolor{black}{where $g_{\it u}$ and $g_{\it l}$ are the statistical weights of the upper and lower states, being 4 and 2 respectively \citep{Goldsmith2012C}. The black line Figure~\ref{Fig:Excitation_Plots} shows how the proportion of \Cplus\ in the upper state therefore varies with $T_{\it ex}$.}

\textcolor{black}{As shown in Equations 5 and 6 of \citet{Gerin2015A}}, the \CII\ integrated line opacity, $\int \tau_{\rm [CII]}\ dv$, in the diffuse ISM is given by:

\begin{equation}
\int \tau_{\rm [CII]}\ dv\ {\rm (km\,s^{-1})} = \frac{N_{{\rm C^+\!,}\,l}}{1.4 \times 10^{17}}
\label{Equation:CII_Tau}
\end{equation}
 
 \noindent where $N_{{\rm C^+\!,}\,l}$ is the column density of \Cplus\ that is in the fine-structure ground state, in cm$^{-2}$.

As $T_{\it ex}$ increases, so too will the column density of \Cplus\ associated with a given integrated line opacity, $\int \tau_{\rm [CII]}\ dv$ (due to the increasing fraction of the \Cplus\ column not contribut\textcolor{black}{ing} to the \CII\ absorption); this effect is plotted with the red line in Figure~\ref{Fig:Excitation_Plots}.

\textcolor{black}{As can be seen in Figure~\ref{Fig:Excitation_Plots}, \CII\ absorption is a near-perfect tracer of \Cplus\ column at $T_{\it ex} < 20\,{\rm K}$, where the vast majority of \Cplus\ is in the ground state. Indeed, \CII\ remains a decent tracer of $N_{\rm C^+}$ even up to $T_{\it ex} = 30\,{\rm K}$, where \textgreater\,91\% of \Cplus\ is still in the ground state. Above this point, however, it breaks down as an accurate probe of $N_{\rm C^+}$ column.}
 
\textcolor{black}{Being able to use \CII\ absorption as a tool to trace the bulk carbon column depends upon the excitation temperature being \textless\,30\,K, and ideally \textless\,25\,K. As discussed in \citet{Gerin2015A}, $T_{\it ex} \leq 25 \,{\rm K}$ is expected for \Cplus\ in the diffuse ISM. }

\textcolor{black}{\citet{Guevara2020B} use multi-tracer multi-isotopologue observations of \CII\ absorption for numerous sightlines\footnote{\citet{Guevara2020B} report the absorbing foreground as difficult to model, as it exhibits significant variation over angular separations as small as 30\arcsec.} towards a number of Galactic star-forming regions, with an average excitation temperature of $T_{\it ex} = 26.8\,{\rm K}$.}

\textcolor{black}{In similar observations of \CII\ absorption towards another Milky Way star-forming region, \citet{Kabanovic2022A} use a comparable approach to infer excitation temperatures of $14 < T_{\it ex} < 30 \,{\rm K}$ for the foreground \Cplus, which their model finds to be associated with \HI\ and low-density H$_{2}$. \citet{Kabanovic2022A} also use optical extinction data to estimate the total ISM column along their long of sight, and thence the total expected C column; from this, they infer that the vast majority of the \Cplus\ must be in the ground state, to account for the observed \CII\ absorption.}

\textcolor{black}{\citet{Langer2016B} find strong \CII\ absorption in most of the 10 sightlines they observe towards the inner Galaxy, and were able to specifically ascribe one of the strongest absorption features to the ISM of the local arm. They report that the excitation temperature of the absorbing \Cplus\ must be `considerably lower' than the $40 \lesssim T_{\it ex} \lesssim 50\,{\rm K}$ of the emitting material.}

\textcolor{black}{It is noteworthy that the low excitation temperatures found by \citet{Langer2016B}, \citet{Guevara2020B}, and \citet{Kabanovic2022A} were despite their sightlines all being towards the Galactic plane -- which should, on average, have considerably {\it higher} excitation temperatures than the diffuse ISM at higher Galactic latitudes. This is due to the greater abundance of warm and ionized material at low Galactic latitude \citep{Goldsmith2012C,Ossenkopf2013A,Langer2014A}. Nonetheless, there has been relatively little study of the excitation of \Cplus\ in the general ISM at any Galactic latitude. Future missions (such as those described in Section~\ref{Section:Prospects}\textcolor{black}{)} will be needed to confirm that the bulk \Cplus\ in the diffuse ISM reliably has a low excitation temperature.}

\textcolor{black}{It is also important to note that \citet{Ingalls2011A} use {\it Spitzer} observations of H$_{2}$ rotational transitions to find evidence for excess collisional excitation in high-latitude Galactic cirrus. Given the relatively low $T^\ast = 92\,{\rm K}$ energy separation of the \CII\ transition, such excitation action will increase the excitation temperature of the high-latitude \Cplus. The excitation energy of the H$_{2}$ rotational transitions is greater than that of \CII\ (being \textgreater510\,K), so a larger fraction \Cplus\ will likely be affected than H$_{2}$. However, \citet{Ingalls2011A} find that only $\approx$2\% of H$_{2}$ in their sightlines is rotationally excited, suggesting such collisional excitation only affects a small fraction of the high-latitude ISM. Additionally, \textcolor{black}{i}t seems reasonable to expect that the high-latitude \Cplus\ should, overall, still have a low excitation temperature. With the advent of JWST, and its unparalleled access to IR H$_{2}$ lines, there are good prospects for great improvement in our understanding of the energetics of the diffuse ISM in the coming years.} 
 
With \CII\ absorption providing a viable means to measure \Cplus\ column density, one can then use the \HI\ and H$_{2}$ column densities (the measurement of which is described in Section~\ref{Subsubsection:Carbon_Column}) along the sightline to determine the gas-phase carbon abundance relative to hydrogen. Comparing this to the total (gas phase plus dust phase) ISM carbon abundance, as measured from the atmospheres of Galactic young stars \citep{Lodders2003A}, thereby provides a measure of the carbon depletion.

\needspace{3\baselineskip} \subsection{Identifying Viable Sightlines} \label{Subsection:Identifying_Viable_Sightlines}

We designed a pilot study to test the implementation of the \CII\ absorption method using SOFIA. For this study, we searched for extragalactic background sources with sufficiently bright FIR dust continuum emission, along sightlines that should have sufficiently strong Galactic 158\,\micron\ \CII\ absorption, to be detectable by the upgraded German REceiver for Astronomy at Terahertz frequencies (upGREAT\footnote{GREAT is a development by the MPI für Radioastronomie and cooperation with the MPI für Sonnensystemforschung and the DLR Institut für Planetenforschung.}; \citealp{Heyminck2012A,Risacher2018A}). Later, in Section~\ref{Section:Prospects}, we use the same method for identifying viable sightlines for observation with potential future facilities. 

Because the concept of using FIR absorption lines to measure depletions is novel, the search process for finding candidate sightlines contains some steps and considerations that may not be familiar to astronomers used to measuring abundances and depletions in the UV. We therefore lay out the search process in detail here. Not only is the search process relevant to our SOFIA study (discussed fully in Section~\ref{Section:SOFIA}) -- but we also consider the specific methodology of the sightline identification process to be a useful research output in and of itself, along with the full list of candidate sightlines, both of which we intend to be of use to future investigators. We later apply the method to various proposed future FIR facilities in Section~\ref{Section:Prospects}.

\needspace{3\baselineskip} \subsubsection{Identifying Potential Background Sources} \label{Subsubsection:Background_Sources}

To identify possible sightlines for observation, we started by performing a systematic analysis of every nearby (\textless\,40\,Mpc) galaxy observed using the 160\,\micron\ band of the Photodetector Array Camera and Spectrometer (PACS; \citealp{Poglitsch2010B}) instrument aboard the \hersc\ Space Observatory \citep{Pilbratt2010D}. The surface brightness measured in the \hersc-PACS 160\,\micron\ band provides an accurate indication of the dust continuum available as a background source for \CII\ absorption at 158\,\micron, and the 13--16\arcsec\ resolution of \hersc-PACS  at 160\,\micron\ is a close match to the 15\arcsec\ resolution of SOFIA-upGREAT. For this analysis, we used the \hersc-PACS data presented by \citet{CJRClark2018A}, from the DustPedia\footnote{\url{https://dustpedia.astro.noa.gr/}} survey \citep{Davies2017A}, providing 160\,\micron\ observations of 753 nearby galaxies. 

For each map, we performed a rough background-subtraction by finding the median surface brightness in the map, and subtracting it\footnote{In order to be a viable background source, a galaxy's  surface brightness needs to be so bright that background subtraction is practically unnecessary (given how much brighter the source will be compared to any possible background/foreground). We therefore deemed it unnecessary to perform any more complex background subtraction than this median subtraction.}. For each galaxy, we found the highest surface-brightness area located within 1\arcmin\ of the galaxy's center. To estimate the surface brightness that would be observed, we took the mean \hersc-PACS 160\,\micron\ surface brightness within a circular aperture 20\arcsec\ in diameter. We opted for a 20\arcsec\ aperture, in contrast to the 15\arcsec\ SOFIA beam, in order to make the surface brightness estimate slightly conservative (given the diluting influence of beam side-lobes, for instance), and to make the estimates also applicable to several proposed future FIR facilities (as described in depth in Section~\ref{Section:Prospects}). We discard candidate sightlines where the Signal-to-Noise ratio (S/N) in this aperture is \textless\,10; this left 432 sightlines.

\needspace{3\baselineskip} \subsubsection{Estimating Carbon Column Density} \label{Subsubsection:Carbon_Column}

Measuring the \Cplus\ abundance along a given sightline requires not only that the continuum be bright enough to be detected, but also that there is enough Galactic \Cplus\ along that sightline to create a \CII\ absorption feature that is deep enough to be detected as well. Relatively weak absorption features can potentially be detected in modest amounts of integration time, as long as the continuum is sufficiently bright (and the instrumental baseline sufficiently stable). 

To estimate the expected carbon column density along each candidate sightline, we first measured the column densities of atomic and molecular hydrogen. To evaluate the atomic hydrogen column density, $N_{\rm HI}$, we used the \HI\ column density maps provided by the H{\sc i}4PI 21\,cm survey \citep{Bekhti2016B}. To evaluate the molecular hydrogen component, we used the all-sky Galactic CO(1-0) and CO(2-1) maps of \citet{Planck2013XIII}\footnote{Specifically, we use their Type II maps.}. For sightlines where CO(1-0) is detected in the \citet{Planck2013XIII} map, we calculated $N_{\rm H_{2}}$, the column density of H$_{2}$, using the standard prescription:

\begin{equation}
N_{\rm H_{2}} = 6.25 \times 10^{19}\ I_{\rm CO(1-0)}\ \alpha_{\rm CO}
\label{Equation:H2_Column}
\end{equation}

\noindent assuming the standard galactic CO-to-H$_{2}$ conversion of $\alpha_{\rm CO} = 3.2\,{\rm M_{\odot}\,pc^{-2}\,K^{-1}\,km^{-1}\,s}$ \citep{Obreschkow2009A,Bolatto2013B}. 

For sightlines where CO(1-0) is not detected in the \citet{Planck2013XIII} map, we instead used their CO(2-1) map, and applied a $\frac{I_{\rm CO(2-1)}}{I_{\rm CO(1-0)}}$ line ratio of $r_{2:1} = 0.7$ \citep{Leroy2013B,Saintonge2017A}, to calculate the H$_{2}$ column density. 

For sightlines where there was insufficient CO(2-1) for \citet{Planck2013XIII} to provide an intensity, we assumed that there was therefore negligible molecular gas present. The \citet{Planck2013XIII} CO(2-1) maps have Root Mean Squared (RMS) noise of $0.12\,{\rm K\,km\,s^{-1}}$, corresponding to a molecular gas column density of $3.4 \times 10^{19}\,{\rm cm^{-2}}$. For comparison, all of our candidate sightlines have \HI\ column densities $>4.2 \times 10^{20}\,{\rm cm^{-2}}$. We can therefore rule out the molecular gas being a considerable fraction of the total column for sightlines where CO(2-1) is undetected by \citet{Planck2013XIII}. This is doubly true given that sightlines with undetected CO will of course tend to be in parts of the sky with lower overall gas column density, and therefore where molecular gas fractions should be lowest anyway.

Having found the total \HI\ plus H$_{2}$ column densities along our candidate sightlines, we calculated the expected column densities of carbon (the total, in both the gas and dust phases) by applying a Milky Way carbon abundance of $12 + {\rm log}_{10} [ \frac{C}{H}] = 8.47$ \citep{Lodders2003A,Jenkins2009B}.

Equation~\ref{Equation:CII_Tau} only applies to \CII\ in the diffuse ISM \textcolor{black}{(as per Section~\ref{Subsection:Absorption_and_Excitation})}. To check if our candidate sightlines are indeed sampling the diffuse ISM, we calculated the molecular gas fraction, $f_{\it mol}$, for each, according to:

\begin{equation}
f_{\it mol} = \frac{\Sigma_{\rm H_{2}}}{\Sigma_{\rm HI} + \Sigma_{\rm H_{2}}}
\label{Equation:Mol_Frac}
\end{equation}

\noindent where $\Sigma_{\rm H_{2}}$ and $\Sigma_{\rm HI}$ are the mass surface densities of \HI\ and $\rm H_{2}$, respectively.
  
We found $f_{\it mol}<0.2$ for 95\%\ of our candidate sightlines, and $f_{\it mol}<0.1$ for 71\%\ of them. This gives us good confidence that these are sampling the diffuse ISM, and that therefore Equation~\ref{Equation:CII_Tau} is applicable. We discarded candidate sightlines with $f_{\it mol}>0.2$ from further consideration, as these sightlines. This left 409 sightlines.

\textcolor{black}{The 16\arcmin\ resolution of H{\sc i}4PI atomic gas data is well-matched to the 15\arcmin\ resolution of the \planck\ molecular gas data. These are both of course at significantly lower resolution than the \hersc-PACS FIR data. However, as we have discarded sightlines with $f_{\it mol}>0.2$, we will primarily be dealing with the diffuse ISM, which tends to vary more smoothly, over larger angular scales, with power spectrum studies of H{\sc i}4PI showing much less power at smaller scales \citep{Choudhuri2017B}. The lower resolution of the gas data will lead us to overestimate $f_{\it mol}$ for some candidate sightlines, and underestimate it for others; but, on average, should still provide a representative sample.}

\needspace{3\baselineskip} \subsubsection{Contaminating \CII\ Emission from the Background Galaxy} \label{Subsubsection:Backround_Contamination}

As further examined in Section~\ref{Subsection:IC342}, \CII\ emission from the background galaxy can contaminate the velocity range where Milky Way \CII\ absorption is expected, compromising our ability to measure any \CII\ absorption. We therefore exclude sightlines where the background galaxy has a radial velocity in the range $-100 < v < 100\,{\rm km s^{-1}}$ (as per the velocities in the DustPedia catalog, as presented in \citealp{Davies2017A} and \citealp{CJRClark2018A}), to ensure there will be no overlap between the \CII\ velocities of the background galaxy and the Milky Way. This left 402 sightlines.

Note that we imposed this criterion based on the results of our SOFIA observations of IC\,342 (see Section~\ref{Subsection:IC342}), which has a velocity of $\approx 35\,{\rm km\,s^{-1}}$. Therefore IC\,342 does not appear in our final catalog of candidate sightlines, despite being targeted by the SOFIA program discussed later in this paper.

\needspace{3\baselineskip} \subsubsection{The Carbon Phase Along our Candidate Sightlines} \label{Subsubsection:Carbon_Phase}

As mentioned at the beginning of Section~\ref{Section:Method}, the vast majority of gas-phase carbon exists as \Cplus, with C$^{0}$ typically being found in a relatively thin shell around molecular clouds, where carbon is then incorporated into CO. Studies of absorption by C$^{0}$ in the UV find C$^{0}$ column densities spanning the range $10^{13.6}$--$10^{14.9}$\,cm$^{-2}$ \citep{Jenkins2011B}. Even if we adopt the upper end of this range, this still corresponds to \textless\,2\%\ of the expected overall gas-phase carbon column for 70\%\ of sightlines -- with every sightline having \textless\,5\%\ of the gas-phase carbon in C$^{0}$. We are therefore happy to treat C$^{0}$ as a negligible portion of the carbon budget for our candidate sightlines.

Ironically, it is challenging to use observations of CO emission to actually estimate the column density of CO molecules along a given sightline -- and hence the fraction of carbon locked up in CO. This is because the optical depth of $^{12}$CO evolves dramatically from the diffuse to dense ISM \citep{Gordon1976B}. The ratio of the intensity of CO emission to the actual column density of CO, $\frac{I_{\rm CO}}{N_{\rm CO}}$, can therefore evolve by almost two orders of magnitude \citep{Shetty2011A}. In the diffuse ISM, $\frac{I_{\rm CO}}{N_{\rm CO}}\approx 1 \times 10^{-15}\,{\rm K\,km\,s^{-1}\,cm^{2}}$; whereas in the dense ISM, $\frac{I_{\rm CO}}{N_{\rm CO}} \approx 3 \times 10^{-17}\,{\rm K\,km\,s^{-1}\,cm^{2}}$ \citep{Burgh2007A,Liszt2010A}. 

Taking the dense ISM $\frac{I_{\rm CO}}{N_{\rm CO}}$ value as a worst-case-scenario, we find that only \textless\,3.9\%\ of the carbon would be locked up in CO, for all our candidate sightlines (with a median fraction of 1.1\%). However, as noted at the end of Section~\ref{Subsubsection:Carbon_Column} above, we have already rejected sightlines with $f_{\it mol}>0.2$, and so should not be sampling the dense ISM. If we instead apply the diffuse ISM $\frac{I_{\rm CO}}{N_{\rm CO}}$ value, then we find that \textless\,0.2\%\ of the carbon along all of our candidate sightlines should be locked up in CO (with a median fraction of 0.04\%). We deem this to be entirely negligible.

To estimate the fraction of the carbon column expected to be in the gas phase (as opposed to the dust phase) for each sightline, we used the hydrogen column density to calculate the predicted depletion of carbon, $\delta_{\rm C}$, using the depletion strength relationship of \citet{Jenkins2009B} (see also \citet{Roman-Duval2022A}), which for carbon they find to be:

\begin{equation}
\delta_{\rm C} = (-0.101 \times (F_{\ast} - 0.803)) - 0.193
\label{Equation:Depletion}
\end{equation}

\noindent where the depletion strength parameter, $F_{\ast}$, is defined as per \citet{Jenkins2009B}, to be a function of hydrogen column density, $N_{\rm H}$ (in cm$^{-2}$), such that:

\begin{equation}
F_{\ast}  = (0.299 \times {\rm log}(N_{\rm H})) - 5.708
\label{Equation:F_star}
\end{equation}

\needspace{3\baselineskip} \subsubsection{Predicting \CII\ Absorption} \label{Subsubsection:Predicting_CII_Absorption}

Having predicted the carbon column density for each of our candidate sightlines, and calculated the fraction of that carbon expected to be in the gas phase, and confirmed that that gas phase carbon should be dominated by \CII, we can now estimate the integrated \CII\ line opacity, $\int \tau_{\rm [CII]}\ dv$, using Equation~\ref{Equation:CII_Tau}. Determining the depth of the corresponding absorption feature requires us to also estimate what the width of the feature will be. 

The velocity width of the \CII\ line is known to correlate with the width of the \HI\ line. Specifically, \CII\ generally tends to have a narrower velocity than \HI, by an average factor of approximately 1.5 \citep{Langer2014A,Mookerjea2016A,DeBlok2016B}, \textcolor{black}{although there is significant variation. However, by discarding candidate sightlines with $f_{\it mol}>0.2$, we have removed the molecular-dominated environs where the correlations is at risk of being weakest.} 

To estimate the width of the \CII\ absorption feature for each candidate sightline, we refer again to \HI\ data of the HI4PI survey, by using their line width maps. These line width maps give the \HI\ velocity width, $\sigma_{\rm HI}$, that would be required to account for all detected emission at a given position, if all of that emission were associated with a single Gaussian component. The HI4PI data release\footnote{\url{https://cade.irap.omp.eu/dokuwiki/doku.php?id=hi4pi}} describes this metric as follows: ``{\it This is a highly robust statistic in the sense that it behaves well in the presence of noise or multiple components. It can easily miss subtleties in the line profile and has some dependence on spectral resolution}''. As such, we consider this a suitable basis for estimating the width of the \CII\ feature; whilst it will not be perfectly accurate, the correspondence between the \HI\ and \CII\ features will, of course, also not be perfect. We therefore estimate the velocity width of the \CII\ feature according to $\sigma_{\rm C^+} = \sigma_{\rm HI} / 1.5$. 

By combining this velocity width information with the predicted integrated \CII\ line opacity, we can estimate the expected \CII\ absorption profile, and thus its peak opacity, $\tau_{\rm [CII]}^{\it (peak)}$. \textcolor{black}{This can be done substituting into a Gaussian absorption profile and the underlying continuum into Equation~\ref{Equation:CII_Tau}. Specifically, a Gaussian absorption profile with mean velocity $\mu_{\it abs}$ in km\,s$^{-1}$, and standard deviation $\sigma_{\it abs}$ in km\,s$^{-1}$, along with absorbing continuum emission with brightness temperature $T_{b}^{\it (cont)}$ in K. Via this substitution, $N_{\rm C^+}$ is related to $\tau_{\rm [CII]}^{\it (peak)}$ by:}

\small
\begin{equation}
N_{\rm C^+} = 
1.4 \times 10^{17} \int -\ln \left( \frac{\exp{\left( \frac{-(v - \mu_{\it abs})^{2}}{2\ \sigma_{\it abs}} -\tau_{\rm [CII]}^{\it (peak)}-1 \right)}}{T_{b}^{\it (cont)}} \right) dv
\label{Equation:CII_Tau_Peak}
\end{equation}
\normalsize

At the velocity where peak absorption occurs, $\tau_{\rm [CII]}^{\it (peak)}$ is related to the observed brightness temperature according to:

\begin{equation}
\tau_{\rm [CII]}^{\it (peak)} = -\ln \left( \frac{T_{b}^{\it (peak)}}{T_{b}^{\it (cont)}} \right)
\label{Equation:Tau_Peak}
\end{equation}

\noindent where and $T_{b}^{\it (peak)}$ is the brightness temperature at the velocity of maximum absorption.

\begin{figure*}
\centering
\includegraphics[width=0.975\textwidth]{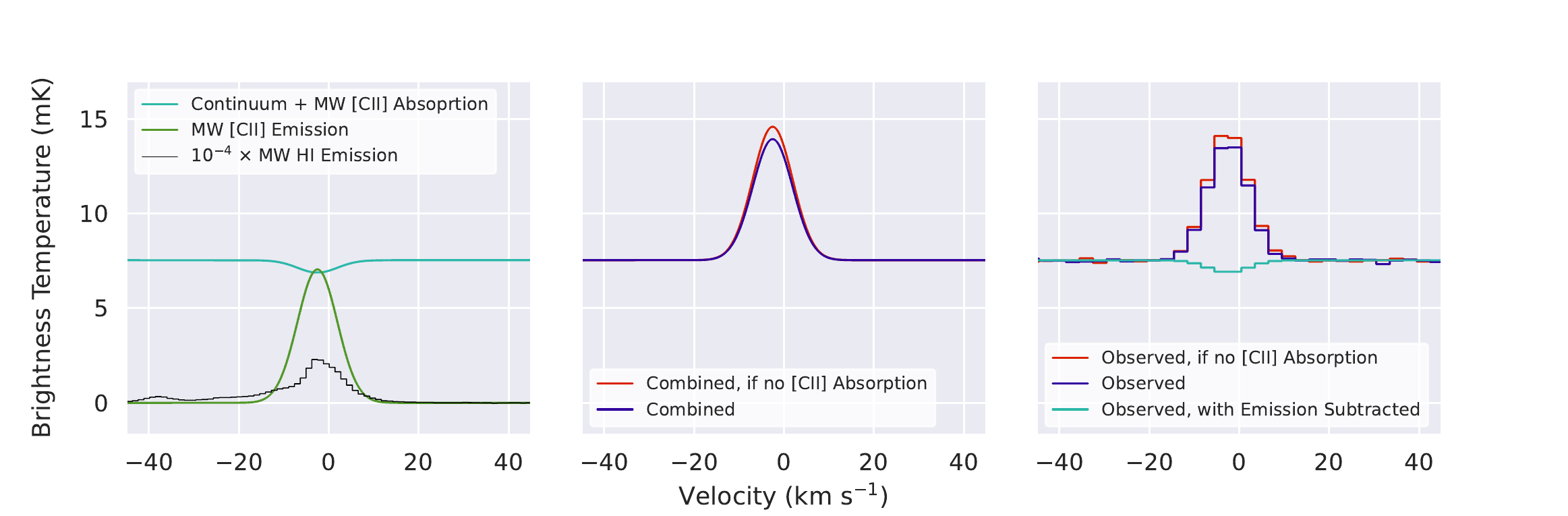}
\caption{Idealized model spectra for the sightline towards background galaxy NGC\,7331. {\it Left:} Predicted model of 158\,\micron\ continuum emission from NGC\,7331 with Milky Way \CII\ absorption; also plotted is the predicted MW \CII\ emission, and the observed Milky Way \HI\ emission (from HI4PI) for reference. {\it Centre:} Predicted combined spectra of background galaxy continuum emission plus Milky Way \CII\ emission, plotted with and without absorption from Milky Way \CII. {\it Right:} Same as centre panel, but with data binned to the $3\,{\rm km\,s^{-1}}$ velocity resolution of FIRSST (see Section~\ref{Subsection:FIRSST_Prospects}), with Gaussian noise added to match level expected from integration to per-channel S/N=10, to simulate observed spectra. Also plotted is the continuum + \CII\ absorption spectrum after the Milky Way \CII\ is subtracted.}
\label{Fig:Example_Spectra}
\end{figure*}

\needspace{3\baselineskip} \subsection{Contamination from Milky Way \CII\ Emission} \label{Subsection:CII_Contamination}

Any observations of the 158\,\micron\ \CII\ spectra of our candidate sightlines not only have the potential of showing Milky Way \CII\ absorption, but also contamination from Milky Way \CII\ {\it emission}. We need to evaluate how severe this contamination could be for each of our sightlines, and to the degree to which it will limit our ability to measure \CII\ absorption. 

It is well established that the surface brightness of the \CII\ line and that of the FIR continuum are related. Previous studies have consistently found that $S_{\rm [CII]} \leq S_{\it FIR} / 200$ \citep{Diaz-Santos2013A,Goicoechea2015C,Sutter2019D}, with $S_{\it FIR}$ defined as:

\begin{equation}
S_{\it FIR} = 1.26\ (S_{60} + S_{100})
\label{Equation:FIR}
\end{equation}

\noindent where $S_{60}$ and $S_{100}$ are irradiance per solid angle observed by the InfraRed Astronomical Satellite (IRAS; \citealp{Neugebauer1984}) in the 60\,\micron\ and 100\,\micron\ bands, in ${\rm W\,m^{-2}\,sr^{-1}}$, as per \citet{Helou1988D}\footnote{Converting from surface brightness in ${\rm Jy\,sr^{-1}}$ to irradiance per solid angle in ${\rm W\,m^{-2}\,sr^{-1}}$ is achieved by multiplying the surface brightness in ${\rm Jy\,sr^{-1}}$ by conversion factors of $2.58 \times 10^{-14}$ and $1.00 \times 10^{-14}$, at 60 and 100\,\micron, respectively. Note that these are conversion factors to convert from ${\rm Jy\,sr^{-1}}$, {\it not} ${\rm MJy\,sr^{-1}}$; we give the conversion factors for ${\rm Jy\,sr^{-1}}$ here to preserve consistency with \citet{Helou1988D}.}. We can therefore use this relationship to assess the maximum likely level of Galactic \CII\ emission at a given position. 

Using the IRAS-IRIS maps of \citet{Miville-Deschenes2005A}, we created 1\degr\,$\times$\,1\degr\ cutouts centred on each candidate sightline, at 60 and 100\,\micron. We then measured the median surface brightness in each band, within an annulus with inner and outer radii of 20\arcmin\ and 30\arcmin, respectively. \textcolor{black}{This means the Galactic emission is being sampled over a much larger area than the \hersc-PACS beam, inevitably averaging over some degree of variation in surface brightness. This was necessary partly due to the much poorer resolution of IRAS (4.3\arcmin). However, this larger annulus would be necessary regardless, in order to avoid sampling the extended FIR emission of the discs of the target background galaxies. The target background galaxies have optical semi major axes (using $R_{25}$ values from the HyperLEDA database; \citealp{Makarov2014A}) as large as 15\arcmin; our 20\arcmin--30\arcmin\ annulus is therefore designed to adequately sample the typical level of Galactic FIR emission towards each position, without being biased by the emission from the background source. We opt to use the same size annulus for all sources, for consistency.}

Using the annular surface brightnesses, we calculated $S_{\it FIR}$, then assumed $S_{\rm [CII]} = S_{\it FIR} / 200$, to provide a plausible worst-case scenario for the surface brightness of Galactic \CII\ emission we can expect. We then calculated the corresponding flux density we would observe within our assumed 20\arcsec\ diameter aperture (as per Section~\ref{Subsubsection:Background_Sources}).

The level of \CII\ emission predicted using the $S_{\it FIR}$ relationship is the {\it integrated} \CII\ emission. To fully evaluate the expected level of contamination, we require a width for the emission profile. For this, we assume the same velocity width as assumed for the absorption profile in Section~\ref{Subsubsection:Predicting_CII_Absorption}. The population of \Cplus\ producing \CII\ emission along a given sightline will not necessarily be exactly the same as the population absorbing, but they should be similar, especially in the diffuse ISM we are concerned with.

\textcolor{black}{To provide a basic reasonableness check of these IRAS-IRIS-based estimates of the contaminating \CII\ emission, we compared them to estimates derived from basic radiative transfer theory. Specifically, given our estimated \CII\ optical depth, $\tau_{\rm [CII]}$, for each candidate sightline, the surface brightness of the corresponding \CII\ emission should be given by:}

\begin{equation}
S_{\rm [CII]} = B(\lambda_{\rm [CII]}, T_{\it ex}) \left( 1 - e^{-\tau_{\rm [CII]}} \right)
\label{Equation:CII_RT}
\end{equation}

\noindent \textcolor{black}{where $B(\lambda_{\rm [CII]}, T_{\it ex})$ is the Planck function evaluated at $\lambda_{\rm [CII]}$, the 157.7\,\micron\ \CII\ line wavelength, for an excitation temperature $T_{\it ex}$. By manually testing a range of $T_{\it ex}$ values, we find that the $S_{\rm [CII]}$ estimates provided for our candidate sightlines by this equation match those we derive from the IRAS-IRIS $S_{\it FIR}$, if we assume excitation temperatures in the range $5 < T_{\it ex} < 10$. This agrees with the $T < 10$ observed in \CII\ surveys of the diffuse Milky Way ISM (\citealp{Fixsen1999A,Velusamy2010D,Velusamy2012B}; see \citealp{Goldsmith2012C} for further examples), indicating that our estimates of the contaminating Milky Way \CII\ emission are sensible.} 

With our predicted intensities and velocity widths for the contaminating Milky Way \CII\ emission along each candidate sightline, we can estimate how it compares to the expected levels of absorption. For the vast majority of the candidate sightlines, the Milky Way \CII\ emission is predicted to be stronger than the absorption feature. In the left panel of Figure~\ref{Fig:Example_Spectra}, we illustrate an example of this by plotting the predicted emission and absorption spectra for the sightline towards background galaxy NGC\,7331\footnote{The \CII\ absorption and emission features are assigned a central velocity equal to the weighted mean velocity of the \HI\ spectrum (which accounts for 80\% of the \HI\ along this sightline according to the HI4PI decomposition).}, where the Galactic \CII\ emission is expected to be $\approx$11 times stronger than the absorption.

Fortunately, as indicated by the central and right panels of Figure~\ref{Fig:Example_Spectra}, we can still make accurate measurements of the \CII\ absorption, even when the contaminating Milky Way \CII\ emission is much stronger: it simply requires a precise subtraction of the contaminating Galactic emission. 

\needspace{3\baselineskip} \subsubsection{Accurate Subtraction of Milky Way \CII\ Emission} \label{Subsubsection:CII_Subtraction}

In order to assess the level of accuracy and precision we can reasonably expect for subtraction of the Galactic \CII\ emission, we performed simulations using IRAS data. These simulations are described in detail in Appendix~\ref{AppendixSection:IRAS_Sims}. But to summarise here, we used IRAS-IRIS 100\,\micron\ data to see how well we could estimate the level of ISM emission within a target aperture with diameter of 1 beam Full-Width Half-Maximum (FWHM), by using two sets sky apertures, located on opposite sides of the target sightline. Each set of sky apertures spanned 3 FWHM, with Nyquist sampling. 

We used Gaussian Process Regression (GPR), a form of probabilistic non-parametric interpolation, to predict the emission level in the target aperture using the emission measured in the sky apertures. For each of our target sightlines, we performed this test at 16 simulated target positions in the IRAS data, offset from the true sightline by 1\degr. By comparing the predicted emission level to the actual measured emission level at the 16 simulated target positions around each sightline, we quantified the accuracy of the sky subtraction possible with IRAS data for each sightline. 

With 16 simulated target positions tested around every one of our candidate sightlines, we had a total of over 6000 simulated sky subtractions. By comparing all of these, we found that it is possible to subtract the sky emission in IRAS-IRIS for a given sightline with a median error of only 1.3\%\ using our GPR method (see Figure~\ref{AppendixFig:IRAS_Sim_Pred}). 

IRAS-IRIS 100\,\micron\ data have an angular resolution over 10 times worse than that of either SOFIA, or of any proposed future FIR facility that might observe \CII\ (see Section~\ref{Section:Prospects}). IRAS is therefore sampling variation in the Galactic ISM at much larger scales than what would be sampled by higher-resolution FIR telescopes -- whereas at smaller scales, the level of variation will tend to become correspondingly smaller. To account for this, we performed additional simulations, also described in Appendix~\ref{AppendixSection:IRAS_Sims}. In these simulations, we repeated the emission subtraction simulations described above, but for a wide range of angular scales, from 1 to 20 times the IRAS FWHM. This allowed us to model the relationship between aperture size, and the accuracy of the resulting sky subtraction. \textcolor{black}{This test assumes that the relationship between angular scale and variation will hold true to angular scales below what can be probed by IRAS. However, it has been well-established that \hersc\ observations of Galactic cirrus show scale-invariant, fractal geometry, down to the \textless\,10\arcsec\ resolution of \hersc\,\footnote{Because \hersc\ data lacks sensitivity to emission on angular scales greater than tens of arcminutes, we cannot perform the test from Appendix~\ref{AppendixSection:IRAS_Sims} directly on \hersc\ observations.}\citep{Ossenkopf-Okada2019A,Roy2019A,Robitaille2019F}. We can therefore rely upon structures, and how they vary, being independent down to the angular scales we are interested in}. With these tests, we extrapolate that it should be possible to subtract sky emission with uncertainty of only 0.33\%\ at resolution $\leq$20\arcsec\ (see Figure~\ref{AppendixFig:IRAS_Sim_Trend}). 

At such a high level of accuracy, we would expect to indeed be able to subtract Galactic \CII\ emission with sufficient precision to accurately retrieve \CII\ absorption features that are many times fainter than the emission, in the manner illustrated in Figure~\ref{Fig:Example_Spectra}. Assuming 0.33\%\ uncertainty on the subtraction for a telescope with 20\arcsec\ resolution, we would expect to be able to detect absorption features at S/N\,\textgreater\,5 as long as the strength of the \CII\ absorption feature is at least $5 \times 0.33\% = 1.65\%$ the strength of the Milky Way \CII\ emission; this corresponds to 144 of our candidate sightlines. Even assuming that the 1.65\% threshold for S/N\,\textgreater\,5 detection is optimistic, a threshold of 4\% would still correspond to 66 candidate sightlines -- a threshold exceeded by the majority of all the sightlines observable by the proposed future facilities discussed in Section~\ref{Section:Prospects}. Given the uncertainty on (and scope for improvement of) the accuracy of the \CII\ subtraction possible with proposed future facilities, we only exclude sightlines for which the predicted \CII\ absorption feature strength is \textless1.65\% of the predicted \CII\ emission strength.

Whilst variation in Galactic dust emission structure at 100\,\micron\ will of course not be a perfect proxy for how Galactic \CII\ emission varies, we should nonetheless expect the different tracers of the diffuse ISM to have a close correspondence, especially in terms of their spatial variation \citep{Planck2013XI}. Moreover, we only used  1-dimensional information for our estimation and subtraction of sky emission in the subtractions we performed here. No doubt the fidelity of the subtraction could be refined still further by using 2-dimensional measurements (ie, making sky observations in both the $x$ and $y$ directions).



\begin{table*}
\centering
\caption{Key properties of the sightlines targeted for our SOFIA study. Note that the NGC\,4945 sightline was not observed before the telescope ceased operations.}
\label{Table:Targets}
\begin{tabular}{lrrr}
\toprule \toprule
\multicolumn{1}{c}{} &
\multicolumn{1}{c}{IC\,342} &
\multicolumn{1}{c}{Circinus} & 
\multicolumn{1}{c}{\it NGC\,4945}\\
\cmidrule(lr){2-4}
Sightline Right Ascension (deg, J2000)									& 56.702  & 213.291 & 196.363 \\
Sightline Declination (deg, J2000)										& +68.096 & -65.339 & -49.467 \\
\cmidrule(lr){1-4}
PACS 160\,\micron\ surface brightness (GJy\,sr$^{-1}$) 					& 8.64 & 30.8 & 77.5 \\
Predicted 157.7\,\micron\ continuum $T_{b}$ (K) 							& 0.078 & 0.278 & 0.699 \\
MW \HI\ column (log$_{10}$\,cm$^{-2}$) 								& 21.55 & 21.72 & 21.14 \\
MW H$_{2}$ column (log$_{10}$\,cm$^{-2}$) 								& 20.38 & 20.39 & 19.24 \\
\cmidrule(lr){1-4}
Predicted MW C depletion (dex) 										& -0.19 & -0.19 & -0.17 \\
Predicted MW gas-phase C column (log$_{10}$\,cm$^{-2}$) 					& 17.69 & 17.72 & 17.30 \\
Predicted \CII\ integrated line opacity, $\int \tau_{\rm [CII]}\ dv$ (km\,s$^{-1}$) 	& 5.49 & 7.96 & 1.98 \\
Predicted \CII\ profile FWHM (km\,s$^{-1}$) 								& 22 & 35 & 14 \\
Predicted \CII\ peak opacity, $\tau_{\rm [CII]}^{\it (peak)}$ 					& 0.25 & 0.20 & 0.14 \\
Predicted MW \CII\ emission (relative to peak absorption) 					& 0.92 & 1.63 & 0.30 \\
\cmidrule(lr){1-4}
Specified SOFIA-upGREAT velocity resolution (km\,s$^{-1}$) 				& 5.1 & 3.3 & 4.4 \\
Specified SOFIA-upGREAT continuum S/N ratio							& 12 & 15 & 20 \\
Specified SOFIA-upGREAT on+off integration time (min) 					& 279 & 49 & 11 \\
Observed SOFIA-upGREAT on+off integration time (min) 					& 77 & 49 & {\it Unobserved} \\
\bottomrule
\end{tabular}
\end{table*}


\needspace{3\baselineskip} \subsection{Catalog of Sightlines} \label{Subsection:Catalog}

The catalog of candidate sightlines is provided in Appendix~\ref{AppendixSection:Catalog}, Table~\ref{AppendixTable:Catalog}, and available online in machine-readable format\footnote{\url{https://iopscience.iop.org/article/10.3847/1538-3881/add40f\#ajadd40ft3}}. The catalog provides all the relevant properties of each sightline, such as position, continuum brightness, and measured $N_{H}$ -- along with the predicted carbon properties, such as $N_{\rm C}$, $\tau_{\rm [CII]}^{\it (peak)}$, and $\sigma_{\it CII}$. The catalog also includes the predicted integration times to detect \CII\ absorption for the sightlines with various proposed future facilities, as described in Section~\ref{Section:Prospects}.

\needspace{3\baselineskip} \section{SOFIA Study} \label{Section:SOFIA}

Following the process laid out in Section~\ref{Subsection:Identifying_Viable_Sightlines}, we found that there are only 5 background galaxies in the sky with 158\,\micron\ continuum emission bright enough to be detected in a reasonable amount of time by SOFIA-upGREAT. Due to atmospheric absorption, and SOFIA's relatively warm primary mirror, continuum emission has to be incredibly bright to be detectable by SOFIA-upGREAT in a practical amount of time, requiring surface brightnesses of \textgreater 5 {\large \it {\bf giga}}janskys per steradian. \textcolor{black}{Although sensitive \CII\ observations of many local galaxies were made by the \hersc-PACS spectrometer (eg, \citealp{Sutter2019D}), that instrument only had velocity resolution of $240\,{\rm km\,s^{-1}}$ (given spectral resolution of 1250 at 158\micron), making it it unsuitable for this experiment, due to excessive dilution of any \CII\ absorption features.}

Frustratingly, two of the background sources with sufficiently bright continuum emission for SOFIA-upGREAT, M\,82 and NGC\,253, lie at relatively high Galactic latitudes, and therefore there is too little Milky Way column for there to be a good chance of \CII\ absorption being detectable in \textless\,10 hours of integration time per galaxy with SOFIA-upGREAT. That left 3 sightlines where Galactic \CII\ absorption could be practically detected against the continuum emission of the background source -- IC\,342, NGC\,4945, and the Circinus Galaxy (ESO\,97-G13, hereafter referred to simply as Circinus). We were awarded 11.4\,hrs (5.6\,hrs of integration, plus overhead) of priority 3 time (ie, filler time) in SOFIA Cycle 9 to observe these sightlines, \textcolor{black}{in program N\textsuperscript{o} 09-0030}. Properties of all three sightlines are given in Table~\ref{Table:Targets}. The target S/N ratio for the continuum along each sightline is calculated to be the S/N required to detect the absorption feature at S/N\,$\geq$\,5. The specified velocity resolution for each was designed to maximise ability to detect absorption, given the velocity width predicted for that sightline according to the \HI\ profile.

Ultimately, only Circinus and IC\,342 were observed; NGC4945 was not observed during our allocated cycles, which were the last cycles of SOFIA operations. Circinus was observed for the full 49 minutes of requested integration, whilst IC\,342 was observed for 77 minutes of integration (30\%\ of the requested total). Circinus was observed during a single flight in August 2021, during SOFIA's southern deployment. IC\,342 was observed over the course of two flights in April 2022.

 \needspace{3\baselineskip} \subsection{General Observing Setup} \label{Subsection:General_Observing_Setup}

Our SOFIA-upGREAT \CII\ observations were made using the instrument's Low Frequency Array (LFA). The LFA is a 14 pixel array with two polarizations; each polarization subarray (horizontal and vertical) has 7 pixels -- with a central pixel surrounded by 6 more pixels in a hexagonal pattern, each offset $\sim$\,30\arcsec\ from the central pixel. We observed in dual-polarization mode; because the horizontal and vertical polarization subarrays are co-spatial, we therefore effectively observed in a manner that treated the LFA as a simple 7 pixel array. For our analysis, we only used the central pixel of this array, as the offsets of the 6 surrounding pixels are large enough that they cannot sample the brightest central region of our target galaxies; meaning the 6 surrounding pixels have no prospect of detecting \CII\ absorption. 

Given how bright the background sources need to be for continuum to be detectable by SOFIA-upGREAT, the strength of the predicted \CII\ absorption was expected to be at a similar level to the contaminating Milky Way \CII\ emission for the sightlines targeted by this program (in contrast to many of the candidate sightlines presented in Section~\ref{Section:Method}, where the emission will often be many times stronger than the absorption). We were therefore confident that the sky subtraction provided by the `off' pointings would do an effective first-order job of removing any Galactic \CII\ emission (in the same manner as it subtracts atmospheric sky brightness). 

We used the standard data products delivered by the observatory team, with the exception that no zeroth-order baseline subtraction was applied (as this would subtract the dust continuum). For each sightline, our specified velocity resolution was set to be the narrowest binning within which we could detect the continuum at our target S/N (as per Table~\ref{Table:Targets}) with Nyquist sampling. Main-beam brightness temperatures, $T_{b}$, were converted from observed antenna temperatures, $T_{A}^{\ast}$, using the SOFIA-upGREAT main beam efficiency of $\eta_{\it MB} = 0.97$, according to $T_{b} = T_{A}^{\ast} / \eta_{\it MB}$, as per the SOFIA Cycle 10 Observer's Handbook\footnote{\url{https://www.sofia.usra.edu/sites/default/files/2022-12/oh-cycle10.pdf}}.

We were unable to achieve a detection with our Circinus observations, due to a strongly-varying instrumental baseline. A full description of the Circinus data is given in Appendix~\ref{AppendixSection:Circinus}. Our observations of IC\,342, however, provided good-quality data.

\needspace{3\baselineskip} \subsection{IC\,342} \label{Subsection:IC342}

Based on our experience with Circinus, our IC\,342 observations were conducted in dual beam-switching mode on the advice of our support scientist, in an attempt to provide a more stable baseline. The maximum 5\arcmin\ throw of the chopping mirror used for beam-switching is small enough that the background positions lie within the FIR emitting disc of IC\,342. However, the average surface brightness of IC\,342 at this radius is a factor of $\approx$\,100 lower than at its center, so this should not meaningfully contaminate the quality of the background measurements. Dual beam-switching mode operates with two background positions, located on opposite sides of the source position. Our background positions were located with 5\arcmin\ throw, at position angles of 105\degr\ and 285\degr\ counter-clockwise from the direction of positive right ascension. We used a on-off chopping frequency of 0.625\,Hz.

We expect the Galactic \CII\ emission along the Circinus sightline to be 92\% as strong as \CII\ absorption \textcolor{black}{versus the continuum}. Therefore, as long as the sky subtraction provided by the off position is accurate to within 22\%, then the absorption feature would still be detectable with uncertainty of \textless20\% (ie, with S/N\textgreater5). As our throw positions were again specifically chosen to be at locations where the level of Galactic emission is a good match to the apparent level in the direction of the sightline itself, we should comfortably exceed this requirement. 

\begin{figure}
\centering
\includegraphics[width=0.475\textwidth]{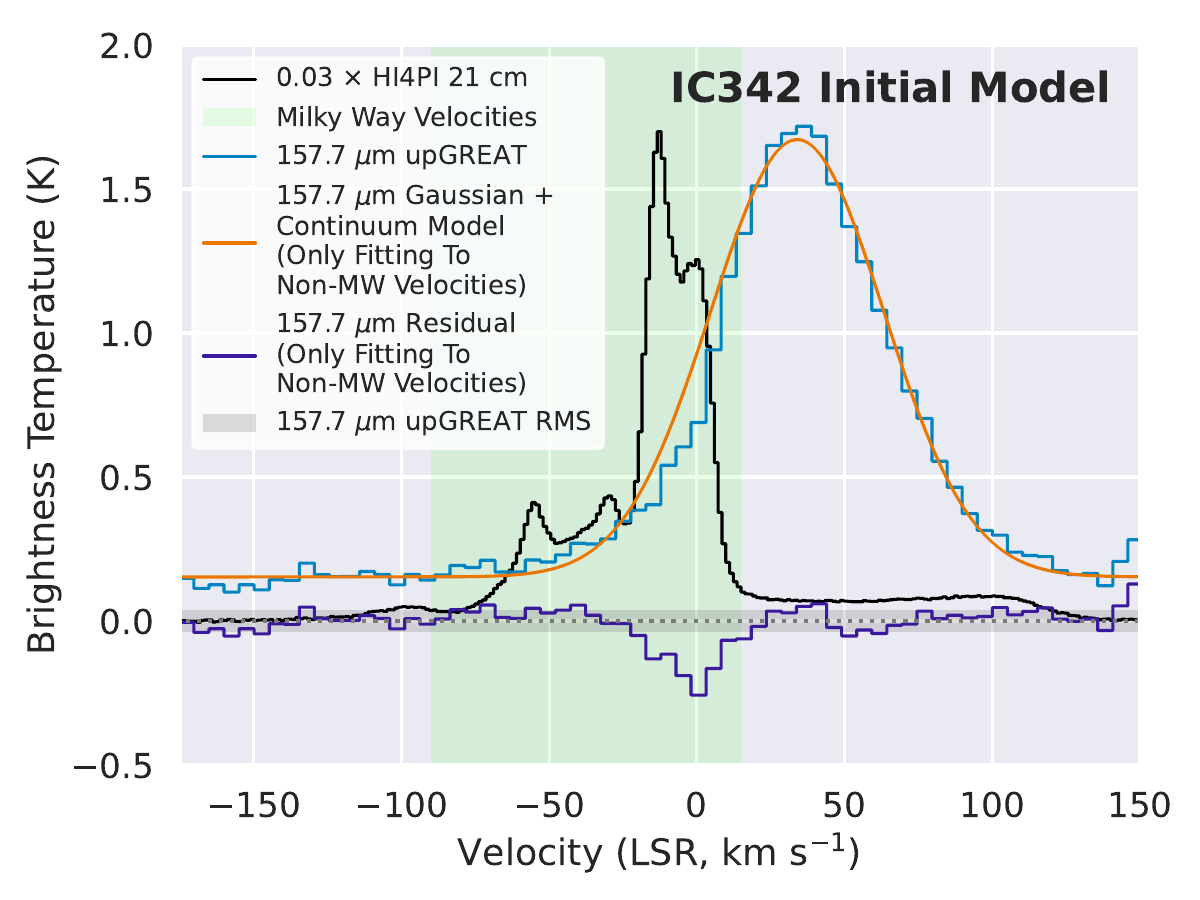}
\caption{SOFIA-upGREAT 157.7\,\micron\ spectrum of the IC\,342 sightline, plotted in blue. Plotted in orange is a model of dust continuum plus Gaussian \CII\ emission from IC\,342 itself, produced by fitting only to data outside the Milky Way velocity range. This model has significant residuals, shown in purple, centered around $v_{\it LSR} \sim 0\,{\rm km\,s^{-1}}$. Also shown for reference, in black, is the HI4PI 21\,cm \HI\ spectrum with $T_{b} \times 0.03$, for ease of comparison with the \CII\ data. The $-90 \gtrsim v_{\it LSR} \gtrsim +15\,{\rm km\,s^{-1}}$ Milky Way \HI\ velocity range is shaded in green. The 21\,cm spectrum exhibits \HI\ emission associated with IC\,342 at $+20 \gtrsim v_{\it LSR} \gtrsim +120\,{\rm km\,s^{-1}}$.}
\label{Fig:SOFIA_Spectra_IC342_1}
\end{figure}

The data has RMS noise of $\varsigma_{\it RMS} = 20\,{\rm mK}$, measured in the velocity ranges $-175 < v_{\it LSR} < -100\,{\rm km\,s^{-1}}$ and  $125 < v_{\it LSR} < 15\,{\rm km\,s^{-1}}$ (chosen to sample only featureless portions of the spectrum).  Our observed spectrum for the IC\,342 sightline is plotted in Figure~\ref{Fig:SOFIA_Spectra_IC342_1}. Also plotted is the HI4PI \HI\ spectrum. The Milky Way \HI\ velocity range is highlighted; there is also a low level of \HI\ emission $+20 \lesssim v_{\it LSR} \lesssim +120\,{\rm km\,s^{-1}}$ that arises from IC\,342.

Our 157.7\,\micron\ spectrum for the IC\,342 sightline contains clear dust continuum emission. However, the spectrum also contains \CII\ emission from IC\,342. This \CII\ emission extends to lower velocities than we had anticipated based on previous SOFIA observations by \citet{Rollig2012B} of two locations in IC\,342. The discrepancy might be accounted for by the fact the \citet{Rollig2012B} observations were much shallower than our own\footnote{The two \citet{Rollig2012B} target positions were observed for only 9.3 and 3.7\,min respectively, versus our 77\,min integration.}, and were of positions offset a few arcseconds from our target sightline. As a result of the \CII\ emission extending to lower velocities than expected, it overlaps the velocity range where we would expect Galactic \CII\ absorption.

Notwithstanding the \CII\ emission, the dust continuum emission from IC\,342 is well-detected and stable. The continuum surface brightness is 152\,mK\footnote{Measured by taking the median brightness temperature within $-175 \gtrsim v_{\it LSR} \gtrsim 150\,{\rm km\,s^{-1}}$, but excluding the Milky Way velocity range of  $-90 \gtrsim v_{\it LSR} \gtrsim +15\,{\rm km\,s^{-1}}$.}, which given the 20\,mK RMS is therefore detected with per-channel S/N\,=\,7.6. This is brighter than the 78\,mK continuum level we estimated based the \hersc-PACS 160\,\micron\ surface brightness. However, as described in Section~\ref{Subsubsection:Background_Sources}, this estimate was intentionally conservative, and was based upon the average 160\,\micron\ surface-brightness within a 20\arcsec\ aperture, which is larger (hence more diluting) than the 15\arcsec\ SOFIA beam.

\needspace{3\baselineskip} \subsubsection{Initial Modeling of the IC\,342 Spectrum} \label{Subsubsection:IC342_Simple_Model}

The contaminating \CII\ emission from IC\,342 complicates the search for Milky Way \CII\ absorption in this data. However, upon visual inspection, we noticed that the \CII\ emission from IC\,342 seems to be asymmetrical. Specifically, it drops off more sharply at lower velocities -- where any absorption from Galactic \CII\ would be expected -- than it does at higher velocities, where the emission profile appears near-perfectly Gaussian.

To test this, we fitted a simple model to the data, consisting of a Gaussian emission profile for the \CII\ emission from IC\,342, along with a flat emission component for the IC\,342 dust continuum. The free parameters were therefore: the dust continuum emission level, $T_{\it cont}$ (in K); the mean velocity of the \CII\ emission Gaussian, $\mu_{\it emit}$ (in km\,s$^{-1}$); the standard deviation of the \CII\ emission Gaussian, $\sigma_{\it emit}$ (in km\,s$^{-1}$); and the amplitude of the \CII\ emission Gaussian, $A_{\it emit}$ (in K). We assumed that the uncertainty on each data point was driven by Gaussian noise, independent in each velocity bin, as charactarized by the RMS noise $\varsigma_{\it RMS}$. 

We performed a maximum-likelihood fit of this model to the data, where we {\it excluded the Milky Way velocity range from the fitting}, so that the effect of any Galactic \CII\ absorption would not affect the modelled profile.

The maximum-likelihood best-fit parameters were: $T_{\it cont} = 0.152\,{\rm K}$, $\mu_{\it emit} = 34.0\,{\rm km\,s^{-1}}$, $\sigma_{\it emit} = 29.3\,{\rm km\,s^{-1}}$, and $A_{\it emit} = 1.52\,{\rm K}$. This model is plotted in Figure~\ref{Fig:SOFIA_Spectra_IC342_1}. Also shown in Figure~\ref{Fig:SOFIA_Spectra_IC342_1} are the residuals between this model and the data. Whilst the model is an excellent fit to the data outside the Milky Way velocity range, there is a very strong negative residual feature centered at $\sim$\,0\,km\,s$^{-1}$ -- corresponding to the velocity range at which the Galactic \HI\ emission peaks, and therefore exactly where we would expect to find any \CII\ absorption from the Milky Way ISM. \textcolor{black}{This negative residual feature is approximately Gaussian, and has a maximum per-channel amplitude of $-$0.26\,K, corresponding to 6.8 times the instrumental RMS. Integrated within the $-50 < v_{\it LSR} < +15\,{\rm km\,s^{-1}}$ velocity range, the negative residual feature has a strength of $-4.49 \pm 0.68\,{\rm K\,km\,s^{-1}}$}

\begin{figure}
\centering
\includegraphics[width=0.475\textwidth]{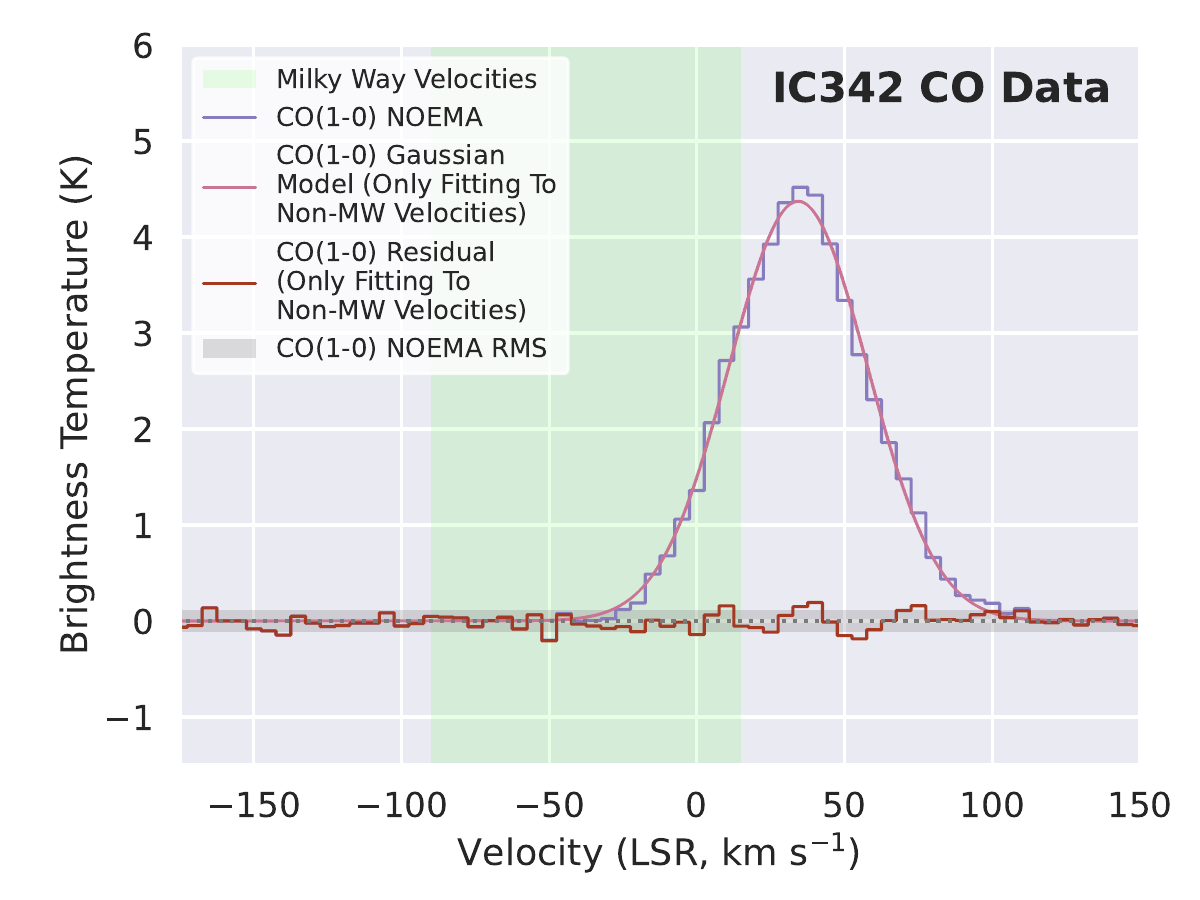}
\caption{\textcolor{black}{Spectrum of CO(1-0) emission from IC\,342 (using data from \citealp{Querejeta2023C}), within the same aperture as our SOFIA-upGREAT 157.7\,\micron\ spectrum, at the same angular resolution. Also plotted is our best-fit Gaussian model to the CO data (produced by fitting only to data outside the Milky Way velocity range), along with its residuals. Unlike with our \CII\ data, the CO emission is well-fit by a Gaussian model, with no significant residuals or evidence of asymmetry.}}
\label{Fig:SOFIA_Spectra_IC342_1.1}
\end{figure}

\needspace{3\baselineskip} \subsubsection{Comparing to IC\,342 CO Emission} \label{Subsubsection:IC342_CO_Model}

\textcolor{black}{The significant asymmetry in the \CII\ emission from IC\,342 is what would be expected to arise from absorption by Milky Way \Cplus\ along the sightline -- assuming the underlying IC\,342 \CII\ emission spectrum is intrinsically Gaussian and symmetric. However, it is also possible that this feature is instead just caused by intrinsic asymmetry in the \CII\ emission from IC\,342, due to velocity structure that happens to align exactly with the velocity at which we would expect Milky Way absorption to occur.}

\textcolor{black}{To examine whether we should expect the \CII\ emission to be intrinsically symmetric, we referred to CO(1-0) observations of IC\,342 presented by \citet{Querejeta2023C}, which combine observations from the NOEMA interferometer and the IRAM 30\,m telescope. The 5\,${\rm m\,s^{-1}}$ velocity resolution of their CO data is a good match to that of our \CII\ data, and the $\leq$4\arcsec\ angular resolution significantly exceeds that of SOFIA.}

\textcolor{black}{We convolved the \citet{Querejeta2023C} CO data cube to match the 15\arcsec\ FWHM of SOFIA at 157.7\,\micron. We then extracted the CO spectra within a 15\arcsec-wide aperture centered at the coordinates of our SOFIA-upGREAT \CII\ spectrum. The resulting CO spectrum should therefore be capturing emission from the same part of IC\,342 as our \CII\ spectrum. The resulting CO spectrum is plotted in Figure~\ref{Fig:SOFIA_Spectra_IC342_1.1}.}

\textcolor{black}{Following the same process as for our \CII\ data, we fit a Gaussian model to the CO spectrum, only fitting to velocities outside the Milky Way velocity range (and omitting the dust continuum component, which will not be detectable in the CO data). The maximum-likelihood best-fit parameters for the CO emission spectrum were: $\mu_{\it emit} = 34.3\,{\rm km\,s^{-1}}$, $\sigma_{\it emit} = 23.4\,{\rm km\,s^{-1}}$, and $A_{\it emit} = 4.37\,{\rm K}$; this model is plotted in Figure~\ref{Fig:SOFIA_Spectra_IC342_1.1}, along with the residuals. Unlike with our \CII\ data, the CO emission is very well fit by a Gaussian model, with no significant residuals or evidence of asymmetry.} 

\textcolor{black}{Molecular gas dominates over atomic in the centers of spiral galaxies, so we should expect CO to be tracing the bulk of the gas \citep{Bigiel2008B,Schruba2011C}. The fact that the CO emission in the center of IC\,342 shows no sign of asymmetry makes it more likely the \CII\ emission, in the exact same aperture, should also be intrinsically \textcolor{black}{symmetric}. This supports the possibility that the negative residual feature in Figure~\ref{Fig:SOFIA_Spectra_IC342_1} is due to absorption from Milky Way \Cplus\ along the line of sight. }

\textcolor{black}{That said, the CO will not be tracing the exact same gas in IC\,342 as the \CII\ emission. The CO should be tracing denser environs than the \CII\ \citep{Wolfire2010A}. This is borne out by the data -- whilst the \CII\ and CO have almost-identical central velocities, their velocity widths differ by 5.9\,${\rm km s^{-1}}$. As such, the well-behaved Gaussian symmetry of the CO emission does not guarantee that the same should be expected of the \CII\ emission.}

\begin{figure}
\centering
\includegraphics[width=0.475\textwidth]{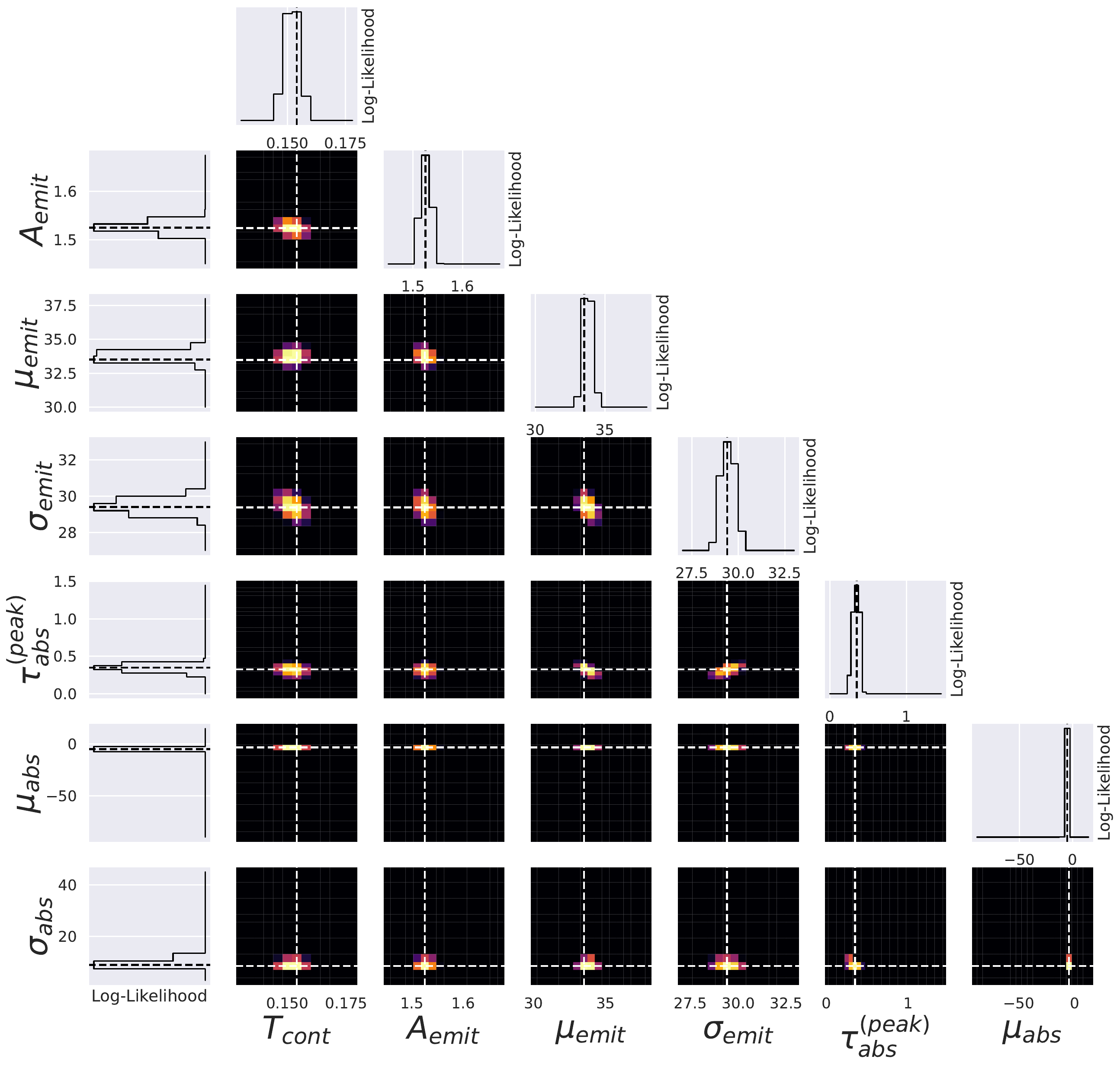}
\caption{Corner plot of the probability distribution of our full model (continuum + \CII\ emission + \CII\ absorption) of the IC\,342 spectrum, as described in Equation~\ref{Equation:Model_Likelihood}, computed via brute-force evaluation of the parameter grid given in Table~\ref{Table:SED_Grid}. The marginalised posterior of each parameter is shown, along with its covariance with each other parameter. The maximum a-posteriori model parameters are indicated with the dashed lines. There is clearly a single, constrained mode in the probability distribution.}
\label{Fig:IC342_Fitting_Corner_Brute}
\end{figure}

\needspace{3\baselineskip} \subsubsection{Robust Modeling of the IC\,342 Spectrum} \label{Subsubsection:IC342_Robust_Model}

\begin{table}
\centering
\caption{SED model grid parameter ranges and step sizes, for the brute-force evaluation of our full model (continuum + \CII\ emission + \CII\ absorption).}
\label{Table:SED_Grid}
\begin{tabular}{lrrrr}
\toprule \toprule
\multicolumn{1}{c}{Parameter} &
\multicolumn{1}{c}{Minimum} &
\multicolumn{1}{c}{Maximum} &
\multicolumn{1}{c}{Spacing} &
\multicolumn{1}{c}{Points} \\
\cmidrule(lr){1-5}
$T_{\it cont}$ (K) & 0.130 & 0.182 & 0.004 & 14 \\
$\mu_{\it emit}$ (km\,s$^{-1}$) & 0.30 & 0.38 & 0.05 & 17 \\
$\sigma_{\it emit}$ (km\,s$^{-1}$) & 27.0 & 33.0 & 0.4 & 16 \\
$A_{\it emit}$ (K) & 1.45 & 1.66 & 0.015 & 15 \\
$\mu_{\it abs}$ (km\,s$^{-1}$) & -90 & 15 & 5 & 22 \\
$\sigma_{\it abs}$ (km\,s$^{-1}$) & 3 & 45 & 3 & 15 \\
$\tau_{\it abs}^{\it (peak)}$& 0.00 & 1.50 & 0.05 & 31 \\
\bottomrule
\end{tabular}
\footnotesize
\justify
\end{table}

\begin{figure}
\centering
\includegraphics[width=0.475\textwidth]{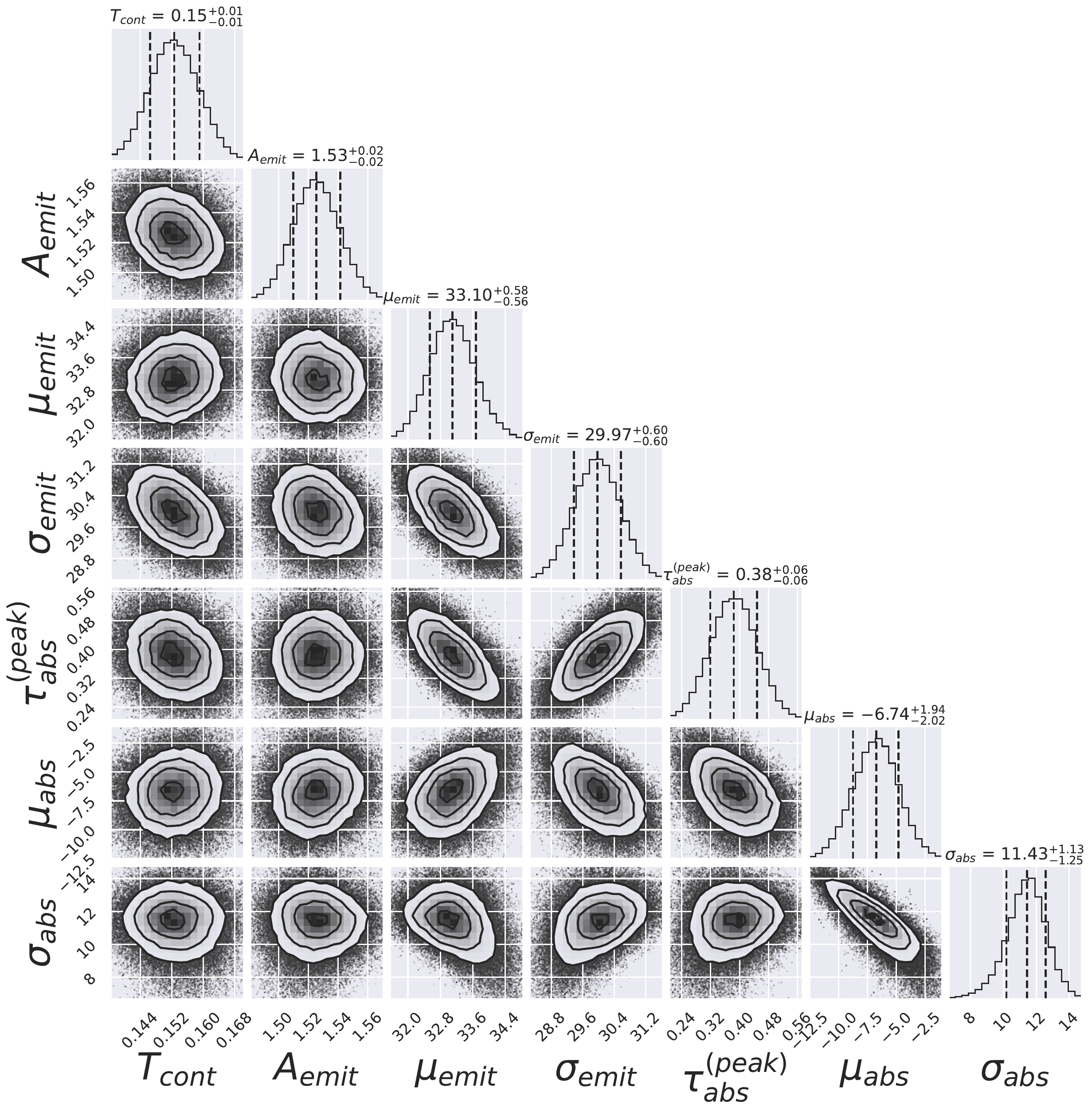}
\caption{Corner plot of the probability distribution of our full model (continuum + \CII\ emission + \CII\ absorption) of the IC\,342 spectrum, from MCMC sampling. The marginalized posterior of each parameter is shown, along with its covariance with the other parameters. For each parameter, the medians of the marginal posterior, along with the 68\textsuperscript{th} percentile credible intervals, are indicated with dashed lines and labeled. The corresponding model spectrum is plotted in Figure~\ref{Fig:SOFIA_Spectra_IC342_2}.}
\label{Fig:IC342_Fitting_Corner_MCMC}
\end{figure}

To further explore the very suggestive negative residual feature in Figure~\ref{Fig:SOFIA_Spectra_IC342_1}, we switch to a more robust modeling approach. We construct a model consisting of a flat dust continuum emission component from IC\,342, a Gaussian \CII\ emission component from IC\,342, and a Gaussian \CII\ absorption component from the Milky Way ISM. This results in 7 parameters; the 4 emission parameters used in the initial model in Section~\ref{Subsubsection:IC342_Simple_Model} above, along with: the mean velocity of the \CII\ absorption Gaussian, $\mu_{\it abs}$ (in km\,s$^{-1}$); the standard deviation of the \CII\ absorption Gaussian, $\sigma_{\it abs}$ (in km\,s$^{-1}$); and the peak opacity of the \CII\ absorption Gaussian, $\tau_{\it abs}^{\it (peak)}$ (ie, the opacity at $\mu_{\it abs}$). We provide a formal description of the model's likelihood function in Appendix~\ref{AppendixSection:Likelihood_Function}

The wide and multi-featured \HI\ profile for the IC\,342 sightline means that there is a broad range of plausible values for the parameters describing any Milky Way \CII\ absorption feature. The absorption parameters will likely also have covariance with the parameters describing the emission from IC\,342. Moreover, there is the risk of there being multiple distinct combinations of parameters that are goods fits to the data. It is impractical to fully sample such a large, 7-dimensional parameter space via Monte Carlo Markov Chain (MCMC) approaches.

\begin{figure}
\centering
\includegraphics[width=0.475\textwidth]{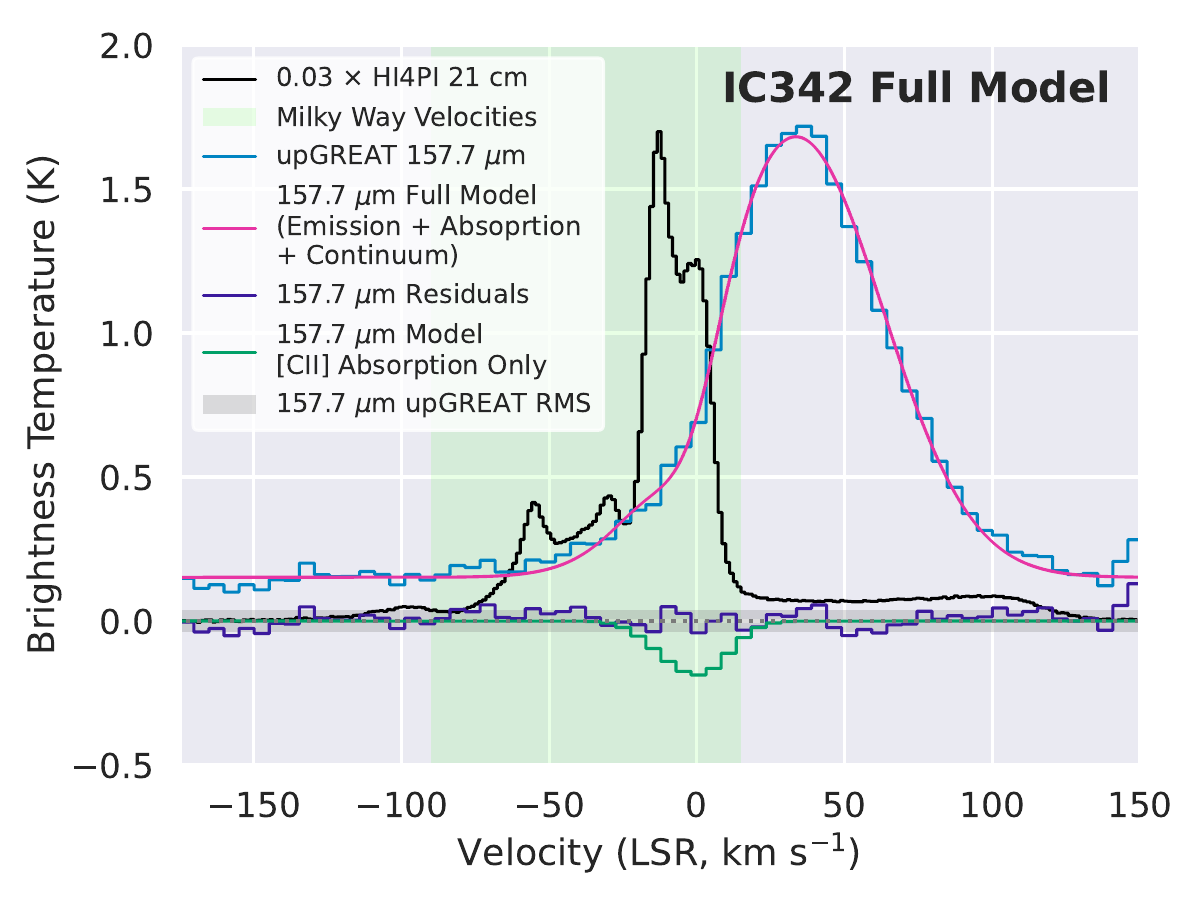}
\caption{SOFIA-upGREAT \CII\ spectrum (blue), HI4PI 21\,cm \HI\ spectrum (black), and Milky Way velocity range (green shading), plotted same as per Figure~\ref{Fig:SOFIA_Spectra_IC342_1}. Also plotted, in pink, is the maximum a-posteriori fit of our full model, which combines dust continuum emission from IC\,342, Gaussian \CII\ emission from IC\,342, and Gaussian \CII\ absorption from the foreground Milky Way ISM, fitted over the full velocity range. Model residuals are shown in purple. Also shown, in teal, is absorption component on its own.}
\label{Fig:SOFIA_Spectra_IC342_2}
\end{figure}

Therefore, we started by performing a full brute-force grid evaluation of this 7-dimensional parameter space. The grid parameter ranges and step sizes are given in Table~\ref{Table:SED_Grid}. For the absorption parameters, we opted to sample very wide ranges, encompassing the full range of values that were even remotely plausible, doing so with relatively narrow grid steps. 

For $\mu_{\it abs}$ (in km\,s$^{-1}$), our grid covers the full Milky Way velocity range. For $\sigma_{\it abs}$, our grid encompasses values as narrow as our velocity binning, up to as wide as 45\,km\,s$^{-1}$ (which corresponds to a FWHM larger than the full Milky Way velocity range). For $\tau_{\it abs}^{\it (peak)}$, our grid range covers values from 0, all the way to 1.5 (which is 6 times greater than the peak opacity we estimated for this sightline; see Table~{\ref{Table:Targets}}). The emission parameters did not require such an extremely wide sampling, as they are reasonably well-constrained by the data in velocity ranges where no absorption is expected; however we still defined grid ranges that extended well above and below the best-fit values found in Section~\ref{Subsubsection:IC342_Simple_Model}. The resulting 7-dimensional parameter grid contains 584\,330\,600 models.

The posterior probability distribution of computed from the brute-force evaluation of our model grid is plotted in Figure~\ref{Fig:IC342_Fitting_Corner_Brute}. As can be seen, there is a single, constrained mode in the posterior. This demonstrates that the posterior does not exhibit multiple complex modes, or other pathologies. Having established this, we could then refine our modeling by exploring the high-probability region of the posterior via MCMC sampling. Specifically, we used the \texttt{emcee} software package for \texttt{Python} \citep{ForemanMackey2013B}. We perform 5000 MCMC steps, with 250 walkers, initialised in a small cluster around the Maximum A-Posteriori (MAP) model identified from our grid search. The first 75\%\ of the steps for each walker are thrown away as burn-in. The resulting sampling of the posterior is shown in Figure~\ref{Fig:IC342_Fitting_Corner_MCMC}.

Figure~\ref{Fig:IC342_Fitting_Corner_MCMC} shows that there is a well-constrained model, featuring a significant absorption component. The median parameter estimates and 68\textsuperscript{th} percentile credible intervals are\footnote{As the marginalized posteriors for all the parameters are effectively symmetrical, we quote a single $\pm$ credible interval for each.}:  $T_{\it cont} = 0.153 \pm 0.006 \,{\rm K}$; $\mu_{\it emit} = 33.1 \pm 0.6\,{\rm km\,s^{-1}}$; $\sigma_{\it emit} = 30.0 \pm 0.6\,{\rm km\,s^{-1}}$; $A_{\it emit} =1.52 \pm 0.02 \,{\rm K}$;  $\mu_{\it abs} = -6.7 \pm 2.0\,{\rm km\,s^{-1}}$; $\sigma_{\it abs} = 11.4 \pm 1.2\,{\rm km\,s^{-1}}$; and $\tau_{\it abs}^{\it (peak)} = 0.38 \pm 0.07$. The corresponding model is plotted in Figure~\ref{Fig:SOFIA_Spectra_IC342_2}. The model is an excellent fit to the data, with residuals that are well-behaved given the RMS.


\needspace{3\baselineskip} \section{Discussion of Potential Detection of \CII\ Absorption} \label{Section:IC342_Discussion}

Our SOFIA-upGREAT data for the IC\,342 sightline is well fit by our model incorporating \CII\ emission and dust emission from IC\,342, and \CII\ absorption from the intervening Milky Way ISM. \textcolor{black}{It is possible that the apparent Milky Way absorption feature in the spectrum is simply due to intrinsic asymmetry in the velocity structure of the \CII\ emission profile of the background galaxy IC\,342 itself. However, it would require us to be {\it exceptionally} unlucky for a circumstantial asymmetry to so closely replicate the properties of the \CII\ absorption feature we were expecting -- especially given the lack any of such asymmetry in the CO spectrum of the exact same region (as per Section~\ref{Subsubsection:IC342_CO_Model}). A model where there are two Gaussian emission components (plus continuum) {\it does} provide an acceptable fit to the data -- although the residuals for this alternative model are significantly worse than those for our model with absorption. Specically, our model with absorption has a $\chi^{2} = 41.2$ in the $-100 < v_{\it LSR} \lesssim 150\,{\rm km\,s^{-1}}$ velocity range -- as compared to $\chi^{2}$ of 67.0 for a model with two emitting Gaussian components and no absorption.}

\textcolor{black}{We do not believe that contamination from Milky Way \CII\ emission could be significantly affecting the observations, or the model. The contribution of Galactic \CII\ emission should be well subtracted from our data by SOFIA-upGREAT dual beam switching (see Section~\ref{Subsection:IC342}). As noted in Table~\ref{Table:Targets}, we predict that the Milky Way \CII\ emission towards this sightline should have brightness corresponding to 92\% of level of absorption expected {\it against the dust continuum emission from IC\,342}. This predicted contamination from Galactic \CII\ emission therefore corresponds to 20\% of the IC\,342 dust continuum level, which equates to 0.03\,K -- much smaller than the \textgreater4\,K amplitude of the absorption we measure against the \CII\ emission from IC\,342. So even if the SOFIA-upGREAT dual beam switching somehow provided a very poor subtraction of the Galactic \CII\ emission, it would still be negligible in amplitude compared to our apparent absorption feature.}

The model has an integrated Galactic \CII\ line opacity of $\int \tau_{\rm [CII]}\ dv = 9.36 \pm 2.38\,{\rm km\,s^{-1}}$ (uncertainty being the 68.3\%\ quantile range of integrated line opacities calculated from the MCMC posterior samples), making it a S/N\,=\,4 measurement. This value of $\int \tau_{\rm [CII]}\ dv$ is 70\% more than our prediction (see Table~\ref{Table:Targets}); this corresponds to a 1.6\,$\sigma$ difference, given the uncertainty. 

Following Equation~\ref{Equation:CII_Tau}, our MAP integrated line opacity corresponds to a total \Cplus\ column density of $(1.31 \pm 0.33) \times 10^{18}\,{\rm cm^{-2}}$. Assuming gas-phase carbon is predominantly in \Cplus\ (see Section~\ref{Subsubsection:Carbon_Phase}), the gas-phase carbon abundance we measure is $12 + {\rm log}_{10} [ \frac{C}{H}] = 8.54^{+0.10}_{-0.13}$.

The \Cplus\ column density we measure therefore exceeds the {\it total} carbon column density we had predicted for this sightline (of $1.14 \times 10^{18}\,{\rm cm^{-2}}$). On face value, this implies that there is no carbon depletion along this sightline; technically we measure carbon `enhancement' (ie, positive depletion) of $\delta_{\rm C} = 0.06^{+0.10}_{-0.13}\,{\rm dex}$. However, the uncertainties on this measurement encompass zero, and indeed extend down to  carbon depletion as strong as $\delta_{\rm C} = -0.07\,{\rm dex}$. In other words, this simply represents a non-detection of carbon depletion for this sightline. Our predicted depletion of $\delta_{\rm C} = -0.19\,{\rm dex}$ lies within 2\,$\sigma$ of our measurement.

We note that our predictions of the expected total carbon column density were based on the assumption the metal abundances of the sightline match the Milky Way average, for which we adopted $12 + {\rm log}_{10} [ \frac{C}{H}] = 8.47$ (as per Section~\ref{Subsubsection:Carbon_Column}). But in practice, ISM metallicity can vary significantly between sightlines. For instance, \citet{Ritchey2023C} find metallicities spanning a range of over 0.4\,dex for a set of 84 sightlines (although most do lie within 0.1\,dex of Solar) -- this is a large enough range that our high carbon column density for the IC\,342 sightline would lie within the expected variation. \textcolor{black}{We also note that different models predict different carbon abundances. The \citet{Hensley2020A} dust model assumes a total interstellar carbon abundance of  $12 + {\rm log}_{10} [ \frac{C}{H}] = 8.52$, which is 0.04\,dex greater than the abundance we adopted; in this case, the range of carbon depletions compatible with our measurement would also be 0.04\,dex greater.}

To gauge how much variation in ISM metallicity we should expect between sightlines, we note that IC\,342 is located at Galactic coordinates of $l=138.17\degr, b=+10.58\degr$, just above the Perseus region of the Galactic plane. The Perseus region contains many of the standard OB star sightlines used for measuring the metallicity of the intervening ISM. \citet{Ritchey2023C} measure the ISM metallicity for 10 sightlines in this area\footnote{Towards the stars HD\,12323, HD\,13268, HD\,13745, HD\,14434, HD\,15137, HD\,23180, HD\,24190, HD\,24398, HD\,24534, and HD\,24912.}. For these sightlines \citet{Ritchey2023C} find metallicities spanning a 0.46\,dex range for non-refractory elements (ie, elements less susceptible to having metallicity determinations affected by depletion), with a standard deviation of 0.13\,dex, and an average per-sightline metallicity uncertainty of 0.09\,dex (see their Table~5). This range of metallicities is greater than the difference between the carbon column density we measure for the IC\,342 sightline, and the Milky Way average. It therefore seems plausible that this sightline is probing a column with somewhat elevated metallicity, giving rise to our larger-than-expected column density of carbon.

As already mentioned, \textcolor{black}{it remains possible that the apparent absorption feature is just due to an intrinsic asymmetry in the \CII\ emission profile of IC\,342, that happens, by conspicuous misfortune, to closely match our expected absorption feature.} We therefore characterise this as a \textcolor{black}{possible detection of Galactic \CII\ absorption. Regardless}, this result highlights that using \CII\ absorption to probe the Milky Way's ISM is a practical technique in general -- even in the disadvantageous situation where \CII\ emission from the background galaxy contaminates the data. 

\needspace{3\baselineskip} \subsection{Lessons Learned from the SOFIA Study} \label{Subsection:Lessons_Learned}

Our experience with this SOFIA program has provided insights that should be useful for efforts to use future facilities to measure carbon abundance and depletion via \CII\ absorption (which we consider more fully in Section~\ref{Section:Prospects} below).

The first key lesson is the importance of baseline stability for these kinds of observations. Except for sightlines where the Galactic \HI\ exhibits a narrow single-featured profile, we cannot know {\it exactly} where any \CII\ absorption features will be located, or how wide they will be. Extracting them from the spectra therefore requires a very stable baseline (ideally with any baseline variations ${\ll {\tau_{\rm [CII]}^{\it (peak)}}}$).

The second key lesson is the importance of considering potential sources of contaminating \CII\ emission. We show in Section~\ref{Subsection:CII_Contamination} that contamination from {\it Milky Way} \CII\ emission should consistently be surmountable. However, our SOFIA-upGREAT data for the IC\,342 sightline demonstrate that observers should err on the side of caution when considering the risk of \CII\ emission from the background galaxy contaminating the Milky Way velocity range. We do note, however, that the additional flux provided by the background \CII\ emission from IC\,342 actually makes the Galactic \CII\ absorption {\it easier} to detect, in an absolute sense. This is because a given $\tau_{\it abs}$ now has more photons to absorb than it otherwise would, giving rise to a greater change in observed brightness, increasing the S/N of the absorption. As such, if an observer believes they can accurately constrain the expected \CII\ emission profile of the background galaxy, this might in fact be an advantageous arrangement, from a pure S/N standpoint.

The third key lesson is that for \CII\ absorption to provide strong constraints on carbon depletion, we need to be able to first constrain the metallicity of the ISM along the sightline being probed. For future facilities able to target large numbers of sightlines, it will be possible to use the average Milky Way ISM carbon abundance when probing relations involving many measurements across the sky. Moreover, future optical--NIR integral field spectroscopic projects the like Local Volume Mapper \citep{Drory2024A} will provide direct auroral-line ISM metallicity measurements across huge swaths of sky, both in and above the Galactic plane. This will allow for much more accurate estimates of the ISM metallicity along a given sightline (although the portion of the column from which such metallicities are measured won't necessarily match the integrated column being probed by \CII\ absorption).

\needspace{3\baselineskip} \section{Prospects for \CII\ Absorption with Future Far-Infrared Telescopes} \label{Section:Prospects}

With the decommissioning of SOFIA, there are now no telescopes in operation capable of performing the FIR spectroscopy necessary to observe \CII\ absorption. However, such observations will be possible with several of the FIR concepts proposed for the NASA astrophysics probe mission solicitation, for possible launch in the 2030s, as per the recommendation of the National Academies' 2020 decadal survey of astronomy \& astrophysics (`Astro2020'; \citealp{Astro2020}). Additionally, Astro2020 recommended a FIR flagship mission, to launch after the UV--optical--NIR Habitable Worlds Observatory, potentially in the 2050s.

\textcolor{black}{Future FIR missions will be able to probe abundances and depletions along sightlines rendered inaccessible in the UV, due to high extinction. In extremis, depletions can currently only be measured towards OB stars with visual extinction up to a few magnitudes, even using {\it Hubble}. For instance, the most reddened OB star towards which \citet{Jenkins2009B} were able to measure depletions (for elements other than iron) was V1074 Sco, a sightline with $E(B-V) = 0.61\,{\rm mag}$ of reddening; for a standard Milky Way extinction law of $R_{V} = 3.1$ \citep{Gordon2003B}, this corresponds to $A_{V} = 2.0\,{\rm mag}$. Assuming that the Habitable Worlds Observatory has UV spectroscopic capabilities equivalent to the {\sc Luvoir} Ultraviolet Multi-Object Spectrograph (LUMOS; \citealp{France2017E}) instrument concept, then it will be able to measure depletions for sightlines a factor of $\approx 300$ fainter than the Space Telescope Imaging Spectrograph (STIS; \citealp{Woodgate1998B}) on HST, used by \citet{Jenkins2009B}. This equates to a $\approx 6\,{\rm mag}$ improvement, meaning the Habitable Worlds Observatory could probe $A_{V} \approx 8.0\,{\rm mag}$ sightlines. Nonetheless, large portions of the sky, notably much of the Galactic plane, exhibit with $A_{V} \geq 10\,{\rm mag}$ \citep{Schlafly2011C}, rendering them inaccessible for measuring depletion, even by the Habitable Worlds Observatory. In the FIR, however, high $A_{V}$ makes \CII\ absorption {\it easier} to detect, as the high dust column does not attenuate the background source, whilst also increasing the absorbing column of \Cplus\ (for a given volume density). This makes future FIR missions highly complimentary to the Habitable Worlds Observatory for studying carbon abundance and depletion.}

\begin{figure*}
\centering
\includegraphics[width=0.75\textwidth]{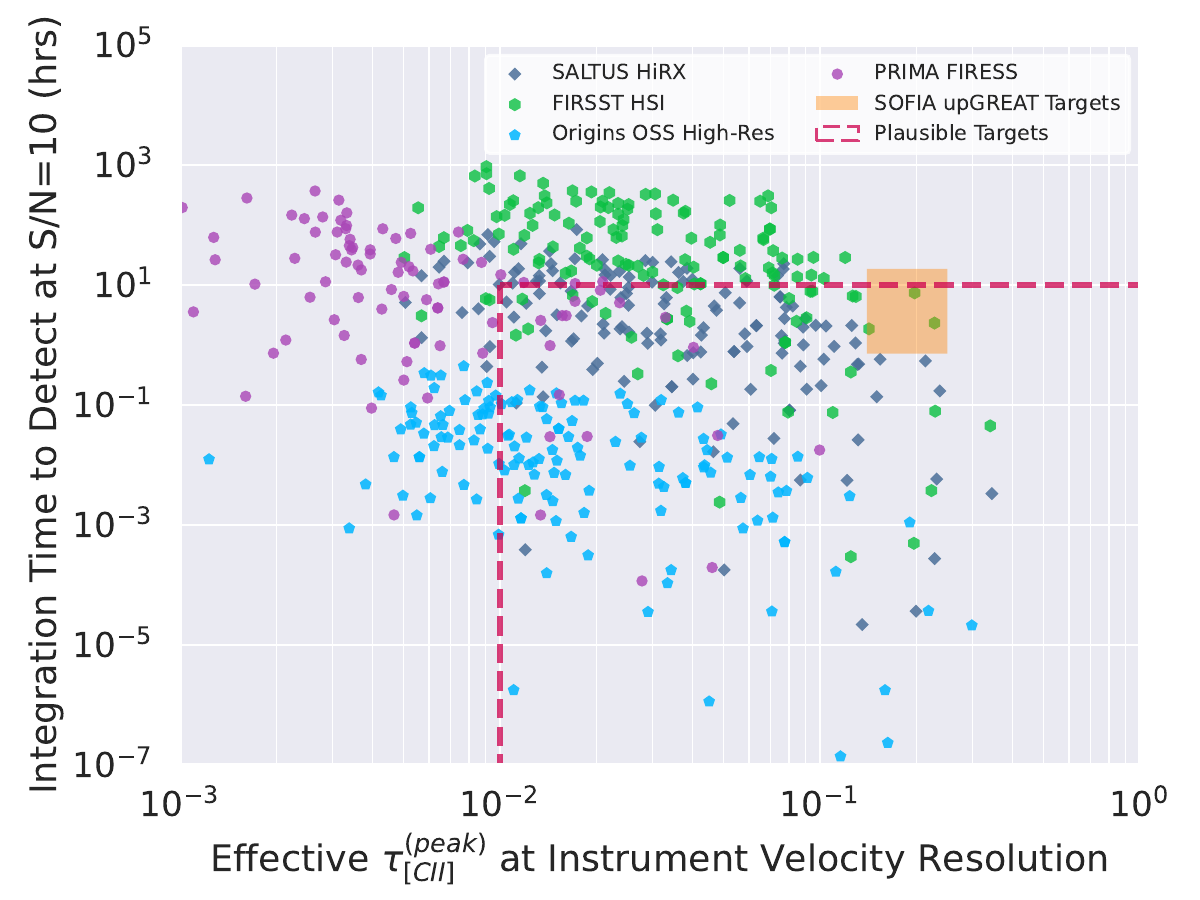}
\caption{Plot of predicted peak \CII\ absorption opacity, $\tau_{\rm [CII]}^{\it (peak)}$, that would be observed at instrumental velocity resolution, against the integration time required to detect that absorption at S/N\,=\,10. Points are plotted for FIRSST, PRIMA, SALTUS, and {\it Origins} proposed FIR telescopes. For each telescope, we plot values corresponding to the 402 diffuse-ISM candidate sightlines identified in Section~\ref{Subsection:Identifying_Viable_Sightlines}.. Additionally highlighted is the region of parameter space occupied by the 3 sightlines targeted by our SOFIA-upGREAT study; these values correspond to the predicted/notional values given in Table~\ref{Table:Targets} (as both observed sightlines returned null results, due to instrumental or astrophysical artefacts). We mark the parameter space where observations should be most plausible (as discussed in Section~\ref{Section:Prospects}). Integration times incorporate any additional time required or sky observations of off positions either side of the target sightline.}
\label{Fig:Future_Predictions}
\end{figure*}

Here, we apply the method laid out in Section~\ref{Subsection:Identifying_Viable_Sightlines}, for identifying sightlines where \CII\ absorption could be detected by multiple proposed FIR facilities, which we describe individually in the subsections below. To estimate the continuum surface brightness that would be observed by each telescope, we convolved the \hersc-PACS 160\,\micron\ observations to the resolution that would be achieved by each telescope at that wavelength\footnote{Except for {\it Origins}, which would have superior resolution to \hersc-PACS; see Section~\ref{Subsection:Origins_Prospects} for details}. We then measure the mean surface brightness of each sightline's background galaxy, inside an aperture with diameter equal to the instrumental FWHM.

We then used the instrumental sensitivities and spectral resolutions for each telescope's design specifications (given in each telescope's subsection below) to estimate the necessary integration time, and expected $\tau_{\rm [CII]}^{\it (peak)}$, for each sightline. We note that $\tau_{\rm [CII]}^{\it (peak)}$ depends on the velocity resolution of the instrument; at lower velocity resolution, the absorption profile will be diluted, reducing the observed $\tau_{\rm [CII]}^{\it (peak)}$ for a given sightline. Instrumental sensitivities are typically provided in terms of the brightness of a spectral feature that could be detected at S/N\,=\,5 in an integration time of 1\,hr. This can then be used to calculate $t_{\it int}$, the time required to achieve S/N\,=\,5 for a spectral feature of a given brightness, according to:

\begin{equation}
t_{\it int} = 1\,{\rm hr} \times \left( \frac{S_{\it 1hr}}{S_{\it target}} \right)^{2}
\label{Equation:Integration_Time}
\end{equation}

\noindent where $S_{\it 1hr}$ is the brightness of a spectral feature that could be detected to S/N\,=\,5 in 1\,hr as per instrumental specification\footnote{Readers accustomed to working at shorter wavelengths should note here that the noise statistics for FIR instrumentation are Gaussian -- in contrast to the Poissonian statistics applicable in the UV, etc.}, in ${\rm W\,m^{-2}}$; and $S_{\it target}$ is the brightness of the spectral feature of interest (such as a \CII\ absorption feature), in ${\rm W\,m^{-2}}$.

Throughout the rest of this section, we adopt a target S/N of 10, to provide a caution factor in identifying sightlines that are plausibly detectable. As described in Section~\ref{Subsubsection:CII_Subtraction}, detecting \CII\ absorption will also require sky observations, to measure and remove the contribution of Galactic \CII\ emission. For each facility discussed below, we also describe how we will incorporate the need for observations of sky pointings, for measuring and subtracting contamination from Galactic \CII\ emission, as per the observing strategy discussed in Appendix~\ref{AppendixSection:IRAS_Sims}.

We plot $t_{\it int}$ against $\tau_{\rm [CII]}^{\it (peak)}$ for each sightline, for each telescope, in Figure~\ref{Fig:Future_Predictions}. Highlighted in that plot is the region of the parameter space where observations should be the most practical (although observations outside this range may of course also be possible).

We consider practically-observable integration times to be \textless\,10\,hrs (including sky pointings). Observations longer than this will likely represent too great a use of telescope time and resources. Most integrations times in the practical range are much less than 10\,hrs.  Some candidate sightlines have such bright continuum, and/or enough predicted \Cplus\ along the column, that the absorption features are notionally detectable in extremely short integration times, sometimes even $\ll$\,1\,sec. Of course, in this regime, the practicalities of instrumentation mean that significantly more time than this would in fact be required -- rather, these sightlines should be considered as requiring `arbitrarily short' integration time, where total observing time in practice will be dictated by read noise, overheads, etc.

We consider practically-observable peak opacity to be $\tau_{\rm [CII]}^{\it (peak)} > 0.01$. This limit is motivated by the degree of baseline stability that can be expected from a well-calibrated space-based MIR--FIR spectrometer. For instance, the James Webb Space Telescope \citep{Gardner2006C,Rigby2023D} Mid InfraRed Instrument \citep{Wright2023A} Medium Resolution Spectrograph has per-band spectro-photometric repeatability\footnote{Which is a measure of to what degree high S/N continuum is found to vary between repeat observations, therefore serving as a measure of baseline stability.} constrained to better than 0.61\% on average, with all bands being better than 1.8\% (\citealp{Argyriou2023B}; specifically, see their Table~4 and Figure~17). The \hersc\ Space Observatory \citep{Pilbratt2010D} Spectral and Photometric Imaging Receiver \citep{Griffin2010D} Fourier Transform Spectrograph had absolute continuum repeatability of \textless2\% (see Figure~3 of \citealp{Hopwood2014A}, and Section~5 of \citealp{Hopwood2014A}). The \hersc\ Hetrodyne Instrument for the Far-Infrared (HiFi; \citealp{deGraauw2010D}) had average short-term continuum repeatability of 2.1\% (see Table~5.8 of \citealp{Herschel-HiFi2017}), although this had a significant contribution from pointing error; for high S/N HiFi observations, the spectral-element-to-spectral-element RMS usually dominates over any baseline variations (see Figure~5.28 of \citealp{Herschel-HiFi2017}). 

We therefore consider it reasonable to expect that $\tau_{\rm [CII]}^{\it (peak)} > 0.01$ features (ie, peak absorption strength of 1\%) will be detectable with future FIR facilities. Indeed, the proposed PRIMA mission (see Section~\ref{Subsection:PRIMA_Prospects} below) is expected to have baseline stability sufficient for \textless\,1\% features to be detectable in a single spectral channel in high-S/N observations (J. Glenn, {\it priv. comm.}).


\needspace{3\baselineskip} \subsection{PRIMA} \label{Subsection:PRIMA_Prospects}

The PRobe far-Infrared Mission for Astrophysics (PRIMA; \citealp{Bradford2022A}) is a mission concept \textcolor{black}{that has been down-selected as the FIR candidate for} NASA's astrophysics probe solicitation. The mission's main focus would be on efficient low-resolution spectroscopic mapping. However, the design also includes a potential medium-resolution mode, the Far-Infrared Enhanced Survey Spectrometer (FIRESS), with spectroscopic resolution of $R = 4\,400$ at 158\,\micron, providing $68\,{\rm km\,s^{-1}}$ velocity resolution, with sensitivity allowing $2 \times 10^{-19}\,{\rm W m^{-2}}$ spectral features to be detected at S/N\,=\,5 in 1\,hr\footnote{Specifications from PRIMA factsheet: \url{https://prima.ipac.caltech.edu/page/fact-sheet}}. PRIMA would have a 2--3\,m primary mirror; to be conservative, we assume the 20\arcsec\ angular resolution of a 2\,m mirror, when using the \hersc-PACS 160\,\micron\ data to estimate the continuum brightness that PRIMA would observe.

The PRIMA mission is not primarily motivated by making high-resolution spectroscopic measurements, so its spectral resolution is more modest than that of SOFIA-upGREAT, or some other proposed FIR facilities. As a consequence, the $68\,{\rm km\,s^{-1}}$ velocity resolution of PRIMA-FIRESS is 5 times wider than the median predicted FWHM of the \CII\ absorption in our candidate sightlines (with 98\%\ of all candidate sightlines having FWHM\,\textless\,$68\,{\rm km\,s^{-1}}$). This means that \CII\ absorption will be significantly diluted for the vast majority of sightlines. Not only does this mean that more integration is required to achieve a detection, but also means that {\it observed} $\tau_{\rm [CII]}^{\it (peak)}$ is expected to be very low. It also raises the risk of `unresolved saturation', where the set of pairs of column densities and linewidths that yield adequate fits to the absorption is highly degenerate.

The modest velocity resolution of PRIMA-FIRESS is the most significant issue. Many absorption sightlines would be unresolved in velocity space, resulting in a low obtainable $\tau_{\rm [CII]}^{\it (peak)}$, as can be seen in Figure~\ref{Fig:Future_Predictions}. There are 17 candidate sightlines that we find could be detected at S/N\,=\,10 by PRIMA-FIRESS in \textless\,10\,hrs of integration time (with another 4 detectable with longer intergrations), and with $\tau_{\rm [CII]}^{\it (peak)} > 0.01$. Whilst not an especially large number, it would still be enough to almost double the 19 existing measurements of interstellar carbon abundance in the literature, from the UV. Additionally, a benefit of PRIMA-FIRESS is that the slit in Band 4 (which includes the 158\,\micron\ wavelength of \CII) has a length of 13\arcmin; therefore no dedicated sky observations would need to be made, as \textgreater10 FWHM of sky sampling would be provided either side of a target centered in the middle of the slit.


\needspace{3\baselineskip} \subsection{FIRSST} \label{Subsection:FIRSST_Prospects}

The Far-InfraRed Spectroscopy Space Telescope (FIRSST; \citealp{McGuire2023A}) is a mission concept \textcolor{black}{that was developed for} NASA's astrophysics probe solicitation. \textcolor{black}{whilst FIRSST was not down-selected for the final selection round, it nonetheless serves as a useful case for considering the capabilities that future facilities may provide}. As the name suggests, the mission has a spectroscopic focus, to probe astrochemistry in various environments. The FIRSST Heterodyne Spectroscopy Instrument (HSI) would have resolving power of $R > 10^{5}$, providing at least $3\,{\rm km\,s^{-1}}$ velocity resolution, with sensitivity allowing $1.8 \times 10^{-19}\,{\rm W m^{-2}}$ spectral features to be detected at S/N\,=\,5 in 1\,hr\footnote{Specifications from \citet{Cooray2023A} and NASA Cosmic Origins Program Analysis Group meeting 2023: \url{https://cor.gsfc.nasa.gov/copag/meetings/AAS_Jan2023/presentations/Cooray-FIRSST-IRTSIG-AAS2023.pdf}}. FIRSST would have a $\approx$2\,m primary mirror; therefore we assume the corresponding 20\arcsec\ angular resolution when using the \hersc-PACS 160\,\micron\ data to estimate observed continuum brightness. FIRSST-HSI would have two 10-pixel arrays in each band; modulo Nyquist sampling, we therefore we assume that two sky pointings would be required to estimate Galactic \CII\ emission for each sightline.

Thanks to the high spectral resolution and sensitivity of FIRSST-HSI, it would be well suited to observations of \CII\ absorption, as can be seen in Figure~\ref{Fig:Future_Predictions}. We find 71 sightlines that could be detected at S/N\,=\,10 by FIRSST-HSI in \textless\,10\,hrs of integration time. Compared to the 19 UV measurements of neutral ISM carbon abundance and depletion in the existing literature, FIRSST-HSI would therefore provide a major step-change in our ability to probe interstellar carbon.

\needspace{3\baselineskip} \subsection{SALTUS} \label{Subsection:SALTUS_Prospects}

\textcolor{black}{The Single Aperture Large Telescope for Universe Studies (SALTUS; \citealp{Chin2024A}) is a mission concept \textcolor{black}{that was developed for} for NASA's astrophysics probe solicitation, \textcolor{black}{although it was not down-selected for the final selection round}. It would use a deployable primary mirror to achieve a 14\,m primary aperture. It's High Resolution Receiver (HiRX) would provide resolving power of R = 10$^{6}$--10$^{7}$, providing $0.3\,{\rm km\,s^{-1}}$ velocity resolution, with sensitivity allowing $5 \times 10^{-19}\,{\rm W m^{-2}}$ spectral features to be detected at S/N\,=\,5 in 1\,hr. The 14\,m primary mirror would be passively cooled by the sunshield, and would provide 2.8\arcsec\ resolution at 158\,\micron.}

\textcolor{black}{This 2.8\arcsec\ resolution dramatically exceeds the resolution of the \hersc-PACS data we have been using to estimate the continuum surface brightness of the background target background galaxies that would observed by each telescope. Such estimates for SALTUS would therefore be significantly overestimated, due to the greatly reduced beam dilution of SALTUS versus \hersc-PACS. No FIR data exists with sufficient resolution to allow us to reliably determine how much brighter the central surface brightness of the background galaxies would be at the SALTUS resolution. The difference would no doubt vary depending on the structure of the FIR-bright core of each galaxy. To be conservative, we simply use the estimates based on the \hersc-PACS data, with the caveat that the resulting estimated integration times for SALTUS are likely overestimated a factor of several. We additionally assume that two sky pointings, to either side of a target, would be required to constrain the Milky Way \CII\ emission towards the a sightline.}

\textcolor{black}{Even with the added conservativeness of our calculations, we still find that SALTUS-HiRX would excel at measuring \CII\ absorption -- we find that 96 of our candidate sightlines would be plausible targets. This is the most of any of the proposed future facilities we consider.}

\needspace{3\baselineskip} \subsection{{\it Origins} Space Telescope} \label{Subsection:Origins_Prospects}

The {\it Origins} Space Telescope \citep{Leisawitz2018B,Meixner2019F} is a mission concept that was submitted to Astro2020, as one of the New Great Observatories\footnote{\url{https://www.greatobservatories.org/}}. A FIR flagship telescope, on the model of {\it Origins}, was recommended by Astro2020, for launch in the decades following the launch of the Habitable Worlds Observatory. Whilst this mission would not necessarily be {\it Origins} per se, we use the published telescope specifications for {\it Origins} here to evaluate the observability of candidate \CII\ absorption sightlines. 

The high-resolution mode of the {\it Origins} Survey Spectrometer (OSS; \citealp{Bradford2018C}) would have resolving power of $R = 10^{4.5}$ at 157.7\,\micron, providing $10\,{\rm km\,s^{-1}}$ velocity resolution, with sensitivity allowing $4 \times 10^{-21}\,{\rm W m^{-2}}$ spectral features to be detected at S/N\,=\,5 in 1\,hr. We use the 20\arcsec\ slit of the {\it Origins}-OSS high-resolution mode to compute the surface brightness it would observe, based on the \hersc-PACS 160\,\micron\ data. The 5.9\,m primary mirror of {\it Origins} would give resolution of 6.7\arcsec, 1.8 times better than that of \hersc-PACS; we therefore leave the \hersc-PACS maps at their native resolution, to provide conservative estimates of the surface brightnesses that {\it Origins} would observe. The 20\arcsec\ slit of the {\it Origins}-OSS high-resolution mode therefore samples 3 FWHM of sky; as such two sky pointings, one either side of the target pointing, would provide sky sampling that exactly replicates the sky subtraction arrangement discussed in Section~\ref{Subsubsection:CII_Subtraction} and Appendix~\ref{AppendixSection:IRAS_Sims}.

{\it Origins}-OSS would also have an ultra-high-resolution mode. Whilst this would have significantly better velocity resolution than the high-resolution mode (by a factor of 7), it also has significantly poorer sensitivity (by a factor of 37). Moreover, the field-of-view of the ultra-high-resolution mode would be only a single 6.7\arcsec\ beam, making it prohibitively time-expensive to conduct enough off pointings for accurate sky estimation. Fewer than 10 of our candidate sightlines would be detectable by the ultra-high-resolution mode in under 10 hours of integration time; all of these sightlines would be easily detectable with the high-resolution mode

With the {\it Origins}-OSS high-resolution mode, we find that 94 candidate sightlines fall in the plausible-target parameter space, plotted in Figure~\ref{Fig:Future_Predictions}. As can be seen in that figure, the limiting factor for {\it Origins}-OSS is not integration time, but rather the $\tau_{\rm [CII]}^{\it (peak)}$ that would be observed. The $10\,{\rm km\,s^{-1}}$ velocity resolution of {\it Origins}-OSS high-resolution mode is wide enough that 71\%\ of candidate sightlines would be unresolved, therefore diluting their observed $\tau_{\rm [CII]}^{\it (peak)}$. Indeed, the exceptional sensitivity of {\it Origins}, driven by its 6\,m class cooled mirror, means that all 94 sightlines could notionally be observed to S/N\,=\,10 in a combined integration time of under 1 hour. 

\needspace{3\baselineskip} \section{Conclusion} \label{Section:Conclusion}

We have presented the novel method of using 158\,\micron\ \CII\ absorption to measure the abundance and depletion of carbon in the neutral ISM. Making these measurements in the UV is extremely challenging, and is currently impossible in the highest-density environments, or in galaxies beyond the Milky Way.

We produced a catalog of 402 sightlines, using FIR-bright nearby galaxies as background source, predicting the Galactic \CII\ absorption that would be measured towards each. This catalog is being made publicly available, for the use of future observers\footnoteref{Footnorte:Catalog_DOI}.

We conducted a pilot study, using SOFIA-upGREAT, in which we attempted to detect \CII\ absorption along 2 sightlines from this catalog, towards the Circinus and IC\,342 galaxies. We report a \textcolor{black}{possible} detection of \CII\ absorption along the IC\,342 sightline, although it requires disentangling \CII\ emission from IC\,342 itself. \textcolor{black}{Whilst the apparent Galactic \CII\ absorption has very high statistical significance, it is not impossible that the apparent \CII\ absorption feature is instead caused by intrinsic structure in the velocity profile of the \CII\ emission from IC\,342, which happens to be very precisely mimicking expected appearance of Galactic \CII\ absorption. In any event, the result demonstrates the practicality of the basic methodology.}

Assuming that gas-phase carbon is predominated by \Cplus, our measured \CII\ absorption corresponds to a gas-phase carbon abundance of $12 + {\rm log}_{10} [\frac{C}{H}] = 8.54$. This is greater than the average Milky Way ISM carbon abundance. No detection of \CII\ absorption was achieved for the Circinus sightline, which we ascribe to an insufficiently stable instrumental baseline.

We also use our catalog of candidate sightlines to assess the potential for three proposed future FIR facilities -- PRIMA, FIRSST, SALTUS, and {\it Origins} -- to detect \CII\ absorption. We find that FIRSST, SALTUS, and {\it Origins}, and to a lesser extend PRIMA, would indeed be able to detect \CII\ absorption along a large number of sightlines. This would dramatically expand our ability to study the abundance, depletion, and behaviour of arguably the most important metal in the diffuse ISM.


{\footnotesize

\section*{Acknowledgments} \label{Section:Acknowledgments}

This research benefited throughout from the constructive, thoughtful, and friendly input (and research environment) provided by the ISM$*$@ST group\footnote{\url{https://www.ismstar.space/}}, whose help made this a better paper.

Based on observations made with the NASA/DLR Stratospheric Observatory for Infrared Astronomy (SOFIA). SOFIA is jointly operated by the Universities Space Research Association, Inc. (USRA), under NASA contract NNA17BF53C, and the Deutsches SOFIA Institut (DSI) under DLR contract 50 OK 2002 to the University of Stuttgart. Financial support for this work was provided by NASA, through SOFIA award N\textsuperscript{o} 09-0030 issued by USRA.

This research made use of \texttt{Astropy}\footnote{\url{https://www.astropy.org/}}, a community-developed core \texttt{Python} package for Astronomy \citep{astropy2013,astropy2019}. This research made use of \texttt{reproject}\footnote{\url{https://reproject.readthedocs.io}}, an \texttt{Astropy}-affiliated \texttt{Python} package for image reprojection. This research made use of the \texttt{NumPy}\footnote{\url{https://numpy.org/}} \citep{VanDerWalt2011B,Harris2020A}, \texttt{SciPy}\footnote{\url{https://scipy.org/}} \citep{SciPy2001,SciPy2020}, and \texttt{Matplotlib}\footnote{\url{https://matplotlib.org/}} \citep{Hunter2007A} packages for \texttt{Python}. This research made use of the \texttt{scikit-learn}\footnote{\url{https://scikit-learn.org}} \citep{Scikit-Learn2011}, a \texttt{Python} package for machine learning. This research made use of the \texttt{pandas}\footnote{\url{https://pandas.pydata.org/}} data structures package for \texttt{Python} \citep{McKinney2010}. This research made use of \texttt{LMFit}\footnote{\url{https://lmfit.github.io/lmfit-py/index.html}}, a non-linear least-squares minimization and curve-fitting package for \texttt{Python} \citep{Newville2015A}. This research made use of \texttt{corner}\footnote{\url{https://corner.readthedocs.io}}, a \texttt{iPython} package for the display of multidimensional samples \citep{ForemanMackey2016D}. This research made use of \texttt{iPython}, an enhanced interactive \texttt{Python} \citep{Perez2007A}.


This research made use of \texttt{TOPCAT}\footnote{\url{http://www.star.bris.ac.uk/~mbt/topcat/}} \citep{Taylor2005A}, an interactive graphical viewer and editor for tabular data.

This research made use of the SIMBAD database\footnote{\url{https://simbad.u-strasbg.fr/simbad/}} \citep{Wenger2000D}, operated at CDS, Strasbourg, France. This research made use of the {\sc Nasa/ipac} Extragalactic Database\footnote{\url{https://ned.ipac.caltech.edu/}} (NED), operated by the Jet Propulsion Laboratory, California Institute of Technology, under contract with NASA. This research made use of the NASA/IPAC InfraRed Science Archive\footnote{\url{https://irsa.ipac.caltech.edu}} (IRSA), which is funded by NASA and operated by the California Institute of Technology.

%




}



\end{linenumbers}
\appendix
\restartappendixnumbering

\needspace{3\baselineskip} \section{SOFIA-upGREAT Observations of Circinus} \label{AppendixSection:Circinus}

Our Circinus observations were conducted in total power mode. We initially chose to use total power mode for all of our observations; although dual beam-switching mode tends to provide a superior sky subtraction, it has a maximum throw of only 5\arcmin, which would be in the outskirts of FIR-emitting discs of our target galaxies. We used a `off' position 11.8\arcmin\ south-east of the center of Circinus, selected to be beyond the FIR structure of Circinus, and in a region with Galactic cirrus surface-brightness approximately equal to that at Circinus' center. We used an on-off cycle with 30\,sec integrations at each position. 

\begin{figure}
\centering
\includegraphics[width=0.475\textwidth]{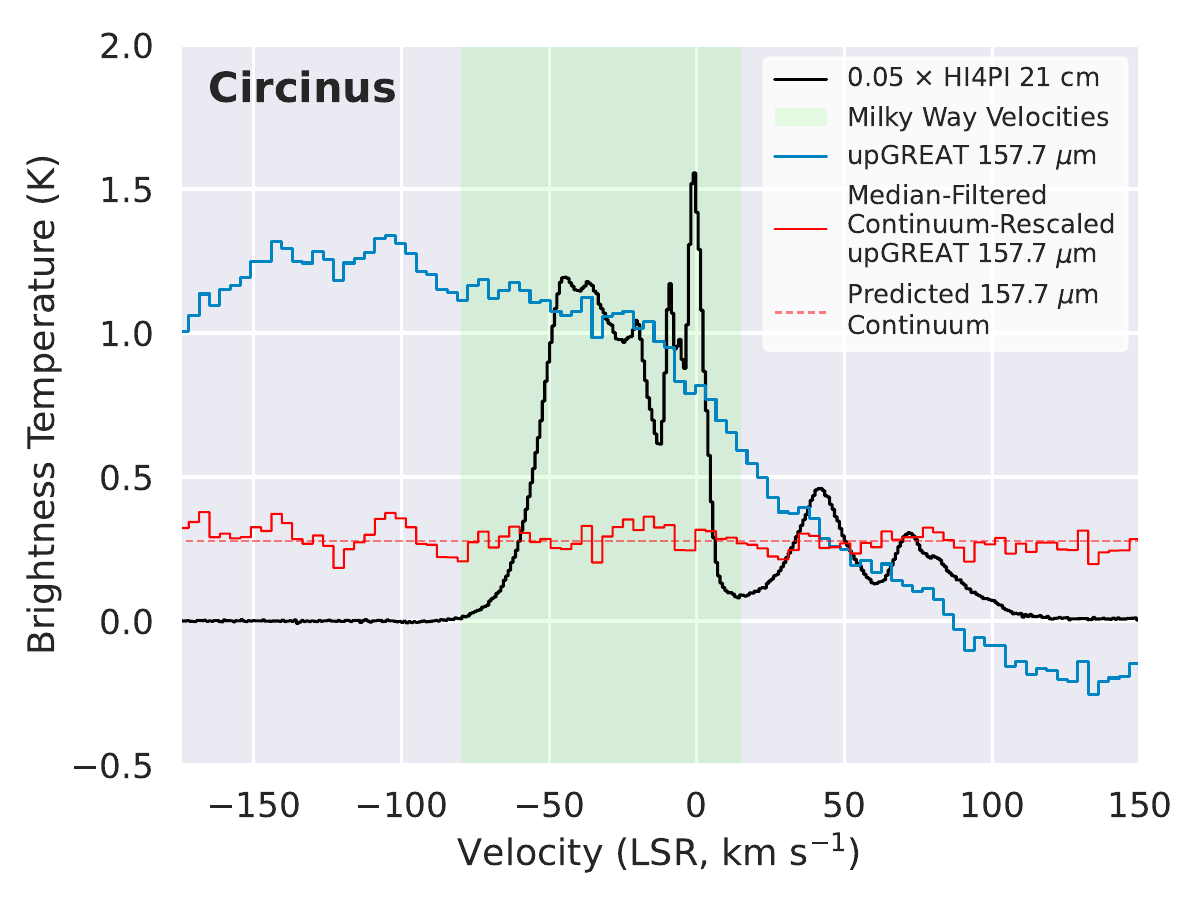}
\caption{SOFIA-upGREAT 157.7\,\micron\ spectrum for the Circinus sightline (in blue), binned to 5\,km\,s$^{-1}$. The SOFIA-upGREAT spectra extend to lower velocities; however, strong atmospheric features dominate the data $< -150\,{\rm km\,s^{-1}}$, so we do not plot those velocity ranges here here. For comparison, we also plot the expected level of dust continuum emission from Circinus, based on measured \hersc-PACS 160\,\micron\ surface brightness. Also plotted (in red) is a version of the \CII\ spectrum that has had a $75\,{\rm km\,s^{-1}}$ median filter applied and subtracted, after which the spectrum was rescaled to the 0.278\,K continuum expected from \hersc-PACS 160\,\micron\ data. Shown for reference (in black) is the HI4PI \HI\ spectrum, with the Milky Way \HI\ velocity range highlighted; the \HI\ spectrum is plotted with $T_{b} \times 0.05$, for ease of comparison with the \CII\ data. Two high-velocity clouds are also present in the \HI\ spectrum towards Circinus (at $\approx\,+40$ and $\approx\,+70\,{\rm km\,s^{-1}}$).}
\label{AppendixFig:SOFIA_Spectra_Circinus}
\end{figure}

Our observed spectrum for the Circinus sightline is plotted in Figure~\ref{AppendixFig:SOFIA_Spectra_Circinus}, displayed at our specified velocity binning of $3.3\, {\rm km\,s^{-1}}$; also plotted is the HI4PI \HI\ spectrum. The Milky Way \HI\ velocity range is highlighted; the two \HI\ features at $\approx\,+40$ and $\approx\,+70\,{\rm km\,s^{-1}}$ are Galactic high-velocity clouds. There is no obvious \CII\ absorption feature. However, the SOFIA-upGREAT spectrum has RMS noise of 41\,mK; this is slightly more than twice the 19\,mK noise we predicted, based on the SOFIA integration time calculator.

The RMS for the Circinus sightline is not helped by the strongly-varying baseline that these observations appear to suffer from. The \textgreater\,1\,K continuum level at $-150 < v_{\it LSR} \lesssim -50\,{\rm km\,s^{-1}}$ is far in excess of the 0.28\,K continuum expected based on the \hersc-PACS 160\,\micron\ surface brightness (also plotted for comparison). Our observatory support scientist suggested that the negative brightness temperatures at $>50\,{\rm km\,s^{-1}}$ are possibly an instrumental artefact arising from poorer instrumental sensitivity at that bluest edge of the passband. 

The other 6 pixels of SOFIA-upGREAT, each offset $\approx$\,30\arcsec\ from the central pixel, do {\it not} show the bright \textgreater\,1\,K continuum we see in the central pixel, suggesting that this may indeed be dust continuum emission from Circinus, as opposed to an instrumental artefact. The absence of this emission feature from the 6 off-center pixels, along with the feature's flat, broad profile, means that we can be confident that this emission is not associated with Galactic \CII\ emission. As per Table~\ref{AppendixTable:Catalog}, we expect the Galactic \CII\ emission along the Circinus sightline to be 1.62 times brighter than the \CII\ absorption. Therefore, as long as the sky subtraction provided by the off position is accurate to within 12.3\%, then the absorption feature would still be detectable with uncertainty of \textless20\% (ie, with S/N\textgreater5). 

In the \hersc\ data for Circinus, the surface brightness of Galactic dust emission within a 10\arcmin--15\arcmin\ annulus around the target (ie, approximately corresponding to the 11.8\arcmin\ throw of the off position) has a standard deviation of 9.6\% (relative to the mean flux in the annulus, which incorporates the absolute flux level based on all-sky data). Assuming variation in dust surface brightness should generally trace variation in diffuse \CII\ surface brightness, then this 9.6\%\ represents the sky-subtraction error we would expect from a randomly-placed off position with 11.8\arcmin\ throw. However, we specifically selected our off position to be as representative as possible of the apparent level of Galactic emission in the direction of the sightline itself, so the accuracy of our foreground subtraction should significantly exceed this. We therefore are confident that the accuracy of the subtraction of Milky Way \CII\ emission provided by the off position will be comfortably better than the 12.3\% required for good measurement of \CII\ absorption

It is not inconceivable that the high continuum brightness temperature we observe in our SOFIA-upGREAT spectrum is due to variability over the decade since Circinus was observed by \hersc-PACS. Circinus is known to host an obscured Active Galactic Nucleus (AGN) with a dusty circumnuclear disc \citep{Tristram2022A} -- and it is known that the luminosity of the hot dust in AGN can vary considerably over these timescales (a fact taken advantage of by dust reverberation mapping; \citealp{Vazquez2015B,Yang2019A}).

The Milky Way \HI\ emission along the Circinus sightline displays a highly structured, multi-component profile. As a result, it is not possible to say with confidence exactly where any \CII\ absorption feature(s) would be expected to lie in this spectrum -- other than {\it somewhere} in the $\sim\,100\,{\rm km\,s^{-1}}$ wide Galactic velocity range. This makes it impractical to attempt complex baseline modeling, in an attempt to locate any weak \CII\ absorption features that may currently be hidden by the high degree of baseline variation -- as we do not know what velocity ranges can be treated as devoid of \CII. Therefore the residuals from any complex baseline modeling would be of similar magnitude any any apparent \CII\ absorption features it generated.

We do, however, perform a simple baseline rectification. We first applied a $75\,{\rm km\,s^{-1}}$ median filter, which we then subtracted from the spectrum. $75\,{\rm km\,s^{-1}}$ is almost as wide as the entire Milky Way \HI\ velocity range, much wider than any of the individual Milky Way \HI\ features, and should be far wider than any potential \CII\ absorption features. This should make a $75\,{\rm km\,s^{-1}}$ median filter a safe choice to remove the smoothly-varying baseline, whilst preserving any more compact features associated with \CII\ absorption. Having subtracted the median filter from the spectrum, we then rescaled the spectrum to be at a brightness temperature of 0.278\,K, to match the continuum predicted from the \hersc-PACS surface brightness. This spectrum is also plotted in Figure~\ref{AppendixFig:SOFIA_Spectra_Circinus}. However, there is still no sign of \CII\ absorption.

Because it is unclear how bright the dust continuum emission detected from Circinus even {\it is}, or if it really is dust continuum that we are seeing in our data (as opposed to some sort of baseline artefact), we are unable to use this non-detection of \CII\ absorption to put any meaningful constraints on the \Cplus\ column along this sightline.

\needspace{3\baselineskip} \section{IC\,342 Spectrum Model Likelihood Function} \label{AppendixSection:Likelihood_Function}

Here we provide a formal description of the likelihood function for the model we use in Section~\ref{Subsubsection:IC342_Robust_Model} to fit our SOFIA-upGREAT spectrum towards IC\,342. A description of the 7 model parameters is provided in Section~\ref{Subsection:IC342}.

Our model's likelihood function assumes that the uncertainty on the data is driven by Gaussian noise, independent in each bin, with constant standard deviation of $\varsigma_{\it RMS}$. Given a spectrum of observed surfaces brightnesses $T_{b}$ (in K), observed at a set of velocities $v$ (in km\,s$^{-1}$), for $n_{v}$ velocity bins, with RMS noise $\varsigma_{\it RMS}$ (in K), for the set of variables $\theta = (T_{\it cont}, \mu_{\it emit}, \sigma_{\it emit}, A_{\it emit}, \mu_{\it abs}, \sigma_{\it abs}, \tau_{\it abs}^{\it (peak)})$, the model's likelihood function $\mathcal{L}$ therefore takes the form:

\begin{multline}
\mathcal{L}(T_{b} |\, v, \varsigma_{\it RMS}, \theta) = \\
\prod^{n_{v}}_{i} \left( \frac{1}{\varsigma_{\it RMS}\sqrt{2 {\rm \pi}}}  \times \exp{\left(\frac{-\left(T_{b_{i}} - T_{\it model_{\, i}}(v_{i},\theta) \right)^{2}}{2\ \varsigma_{\it RMS}^{2}}\right)} \right)
\label{Equation:Model_Likelihood}
\end{multline}

\noindent where for the $v_{i}$, being the $i$\textsuperscript{th} velocity in the set, the brightness temperature arising from a given set of model parameters $\theta$ is given by $T_{\it model_{\, i}}(v_{i},\theta)$, such that:

\begin{equation}
T_{\it model_{\, i}}(v_{i},\theta) = T_{\it cont} + T_{\it emit_{\,i}} + T_{\it abs_{\,i}}
\label{Equation:Predicted_Model}
\end{equation}

\noindent where $T_{\it emit_{\, i}}$ is the \CII\ emission from IC\,342 itself, arising at velocity $v_{i}$, according to:

\begin{equation}
T_{\it emit_{\, i}} = A_{\it emit} \times \exp{\left( \frac{-(v_{i} - \mu_{\it emit})^{2}}{2\ \sigma_{\it emit}} \right)}
\label{Equation:Emission_Model}
\end{equation}

\noindent and where $T_{\it abs_{\, i}}$ is the \CII\ absorption from the intervening Milky Way ISM, arising at velocity $v_{i}$, according to:

\begin{equation}
T_{\it abs_{\, i}} = \exp(-\tau_{\it abs}^{\it (peak)} - 1) \times \exp{\left( \frac{-(v_{i} - \mu_{\it abs})^{2}}{2\ \sigma_{\it abs}} \right)}
\label{Equation:Absorption_Model}
\end{equation}

In total, our parameter grid samples over 584 million points. We apply uniform priors across the entire parameter grid, with one exception: Zero likelihood is assigned to any model which results in an absorption profile where the level of absorption exceeds $2\times\varsigma_{\it RMS}$ anywhere outside the Milky Way velocity range of $-90 \gtrsim v_{\it LSR} \gtrsim +15\,{\rm km\,s^{-1}}$. 

To explain why we impose this constraint, consider a model where there is very strong \CII\ absorption (eg, with $\tau_{\it abs}^{\it (peak)} > 1$) with a broad absorption profile of width $\sigma_{\it abs} = 40\,{\rm km\,s^{-1}}$, centered at a velocity of $\mu_{\it abs} = 10\,{\rm km\,s^{-1}}$ (ie, just inside the upper end of the Milky Way velocity range). Such a model would have high levels of absorption occurring at velocities well above the upper limit of the Milky Way velocity range, in a region where we have strong physical reasons to expect there to be no \CII\ absorption. As long as the \CII\ {\it emission} component of the model was correspondingly strong, the combined model could be an adequate fit to the data, despite being unphysical. Each of these values of the absorption parameters are {\it individually} plausible, just not in this specific combination. Hence we impose the prior that any absorption outside the Milky Way velocity range must have a strength $<2\times\varsigma_{\it RMS}$.

\needspace{3\baselineskip} \section{IRAS Emission-Subtraction Simulations} \label{AppendixSection:IRAS_Sims}

\begin{figure}
\centering
\includegraphics[width=0.475\textwidth]{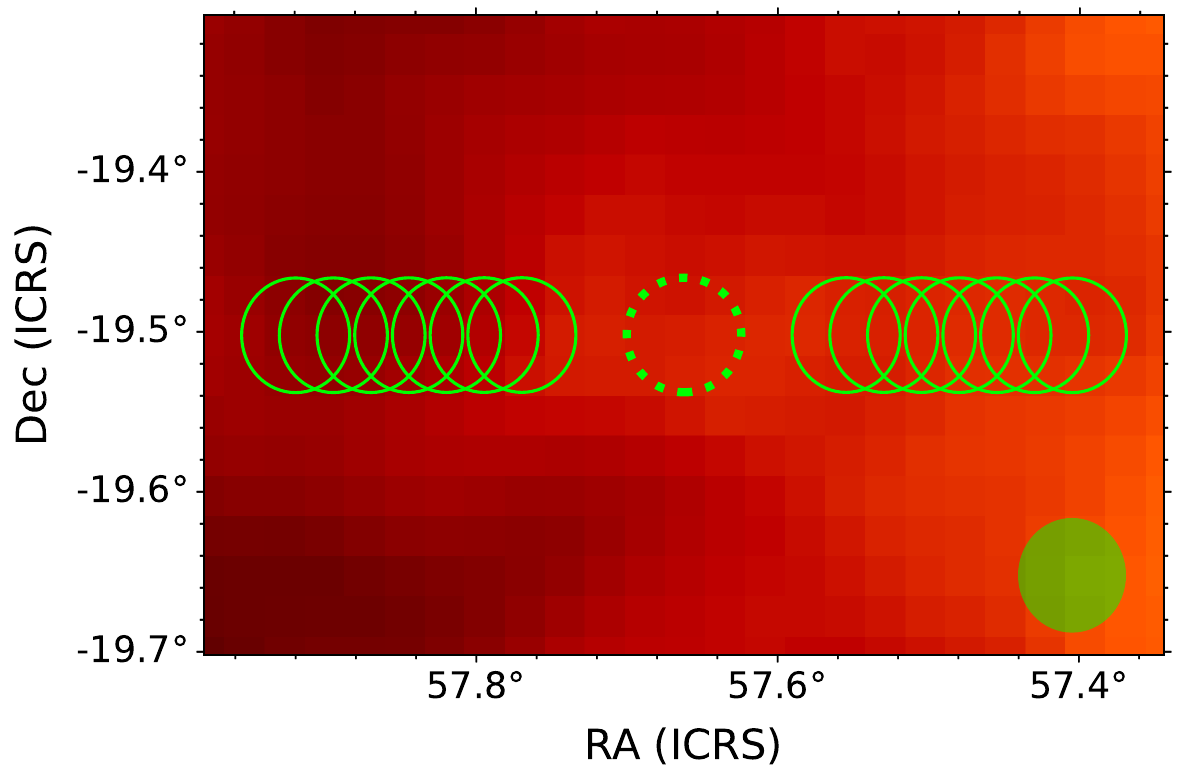}
\caption{Illustration of aperture arrangement used in our IRAS simulations of sky estimation and subtraction, for simulated sightline offset 1\degr\ away from the true sightline for NGC\,1266. Dotted circle indicates the aperture at simulated target position (diameter 1 FWHM) for which we wish to estimate and subtract the emission level. Solid circles indicate the sky apertures, with a set of sky apertures located either side of the target. Each set of sky apertures covers 3 FWHM of sky along the sampling axis. The apertures within each set are spaced with separation of $^{1}/_{3}$ FWHM. The 4.3\arcmin\ FWHM of the IRAS-IRIS 100\,\micron\ beam is illustrated by the filled translucent circle in the lower right corner.}
\label{AppendixFig:IRAS_Sim_Apertures}
\end{figure}

In order to detect and measure the strength of Milky Way \CII\ line absorption seen against the continuum emission bright background galaxies, we need to be able to accurately constrain and subtract the contribution of Milky Way \CII\ emission to the observations.

Here, we describe simulations we have performed to quantify how accurately the emission in a given target location can be predicted, when using measurements of the emission to either side of the target emission. We performed these simulations using the IRAS-IRIS 100\,\micron\ data, as these provide the highest-resolution and most-sensitive maps of the diffuse ISM over large areas. 

For each of the 402 candidate sightlines presented in Section~\ref{Section:Method}, we produced 8 cutouts, $1\degr\,\times\,1\degr$ in size, from the IRAS-IRIS data. Each of these 8 cutouts was located at an offset of $-1\degr$, $0\degr$, or $1\degr$ in both the North--South and East--West directions (excluding the case of there being 0 offset in both dimensions, as this would correspond to no offset). In other words, the offset cutouts were located at the 8 edge positions in a $3\degr\times3\degr$ grid centred on the target sightline. These positions should provide reasonable sampling of the typical variation in the diffuse ISM emission in the area of sky surrounding each candidate sightline.

For each of the 8 offset positions, we measured the average surface brightness within an aperture with diameter of 4.3\arcmin, equal to the FWHM of the IRAS-IRIS 100\,\micron\ beam. This represents the true target surface brightness of this position, to which we will compare our estimates. This target aperture is illustrated by the dotted circle in Figure~\ref{AppendixFig:IRAS_Sim_Apertures}.

We then placed sky apertures to either side of the target aperture, with centres ranging from 1.5--4.5 times the IRAS beam FWHM, with apertures placed every 86\arcsec\ (being $^{1}/_{3}$ of the IRAS-IRIS beam). This is illustrated by the solid-outlined circles in Figure~\ref{AppendixFig:IRAS_Sim_Apertures}. Within each aperture, we measured the average surface brightness. This arrangement therefore provides Nyquist sampling of the variation in the diffuse ISM emission over 3 FWHM of sky, to either side of the target position. 

This system of sky apertures is designed to represent the background sampling that will be practical with various proposed future facilities. For instance, the slit of the high-resolution mode of {\it Origins}-OSS (see Section~\ref{Subsection:Origins_Prospects}) would have a 20\arcsec\ length, corresponding to 3 times the 6.7\arcsec\ FWHM of the PSF at 158\,\micron; we would therefore expect to have 2 background pointings either side of a target when observing with {\it Origins}-OSS. Similarly, the slit for band 4 of PRIMA-FIRESS, which covers the 158\,\micron\ wavelength range, has a length of 13\arcmin, sampling 40 FWHM along its length, assuming the 20\arcsec\ resolution of a 2\,m primary mirror at this wavelength; this would provide very ample background coverage either side of a target source, even in a single pointing. 

\begin{figure}
\centering
\includegraphics[width=0.475\textwidth]{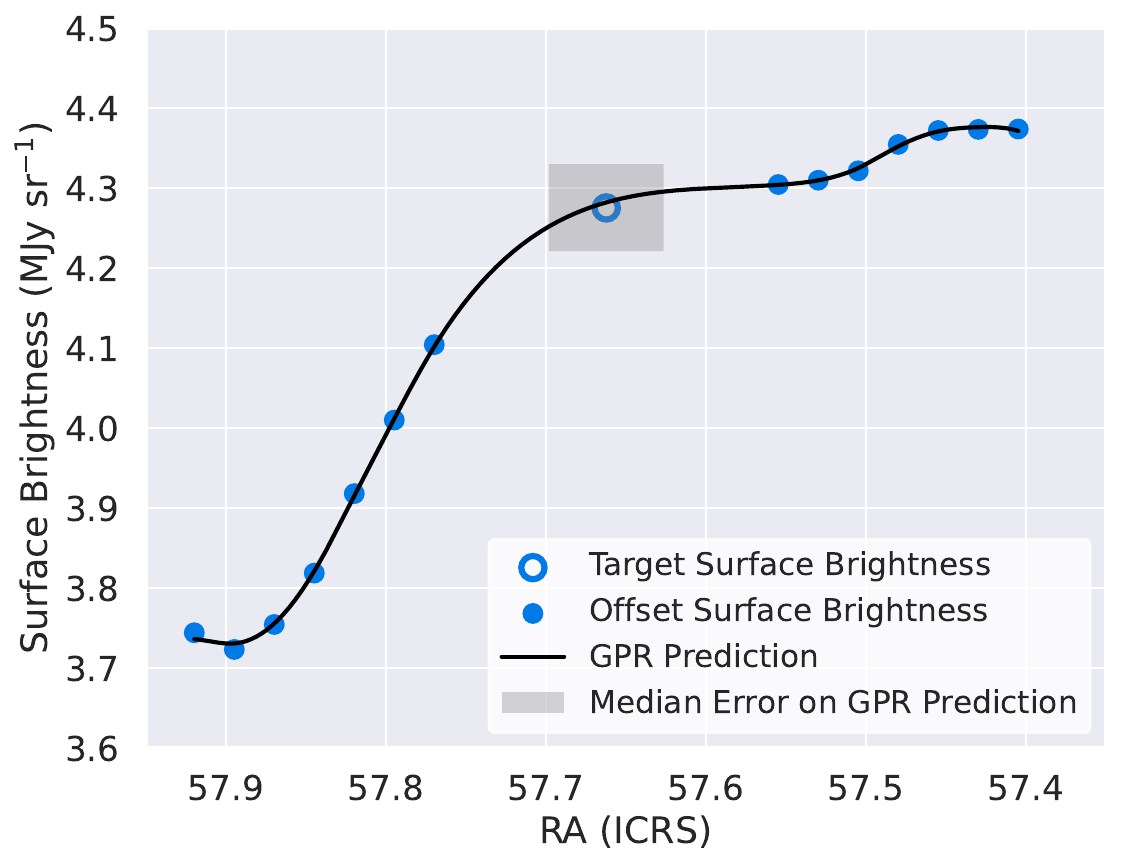}
\caption{Illustration of how we use Gaussian Process Regression (GPR) to estimate and subtract sky emission levels for target sightlines in our simulations. Measured surface brightnesses are for the example sightline shown in Figure~\ref{AppendixFig:IRAS_Sim_Pred}. Vertical height of the shaded region shows the 1.4\% average error on sky estimation that we find across all sightlines. Width of the shaded region corresponds to the 4.3\arcmin\ FWHM diameter of the apertures.}
\label{AppendixFig:IRAS_Sim_Pred}
\end{figure}

We use the surface brightnesses in the sky apertures to predict the sky brightness expected in the target aperture; we can then use the true measured surface brightness in the target aperture to test how accurate the prediction is. We performed the prediction using Gaussian Process Regression (GPR), a form of probabilistic non-parametric modeling, which uses a kernel to identify the characteristic scale over which measurements are covariant. In this instance, we used a M\'atern kernel \citep{Stein1999}. For a fuller description and exploration of Gaussian processes, see \citet{Rasmussen2006} and \citet{Angus2018A}. 

We illustrate the GPR prediction process in Figure~\ref{AppendixFig:IRAS_Sim_Pred}. This shows how can can compare the measured surface brightness at the target position, to the surface brightness predicted by the GPR based on the sky apertures. We thereby compute the error on the prediction. At each of the 8 offset target positions around a given candidate sightline, we repeat this GPR prediction process twice; once with the apertures positioned in the East--West direction, and once with the apertures positioned in the North--South direction. This provides a total of 16 measurements on the error on the surface-brightness prediction in the region of each of our candidate sightlines. 

We sigma-clip the results of each of these 16 measurements, to reject any that at 3$\sigma$ deviant from the average, as such outliers consistently represent regions where interloping bright sources contaminate the target and/or sky apertures. In reality, sky measurements would be targeted to avoid any such interlopers, so their inclusion would make the results unrepresentative. 

For each candidate sightline, after sigma-clipping, we compute the median of the errors in the predicted surface brightness at the offset targets positions. We take this to represent how accurately the diffuse Galactic emission could be quantified and subtracted along that particular sightline. This uncertainty is provided for each candidate sightline in our catalog, as presented in Table~\ref{AppendixTable:Catalog}.

Across all 402 candidate sightlines, we find a the median error on the subtraction is 1.3\%. This number is based on over 6,000 simulated subtractions (with up to 16 simulations for each of the 402 candidate sightlines). We find 1.3\%\ to be impressively accurate, and it bodes well for the prospects of being able to accurately estimate and subtract the contribution of Milky Way \CII\ emission along our candidate \CII\ absorption sightlines -- even when the emission is much stronger than the absorption.

\needspace{3\baselineskip} \subsection{Extrapolating Emission-Subtraction to Smaller Scales} \label{AppendixSubsection:IRAS_Sim_Extrapolation}

However, this 1.3\% accuracy is what we find {\it at the 4.3\arcmin\ resolution of the IRAS-IRIS 100\,\micron\ data}. For the various proposed future FIR facilities discussed in Section~\ref{Section:Prospects}, the angular resolution will be at least as good as 20\arcsec. As noted above, as the magnitude of the variation in the structure of the diffuse ISM falls at smaller angular scales, so too should the errors on the foreground subtraction and estimation.

We therefore should be able to expect even better than 1.3\% accuracy on the subtraction of the Milky Way emission, when operating with the superior resolution of a proposed future FIR facility. To quantify how much better, we performed an additional round of simulations. We defined a range of aperture sizes, evenly spaced logarithmically, from a diameter of 4.3\arcmin\ (1 IRAS-IRIS 100\,\micron\ FWHM) up to 86\arcmin\ (20 IRAS-IRIS 100\,\micron\ FWHM). 

For each aperture scale in the simulation, we degraded the resolution of the data, reprojecting the IRAS cutouts such that the pixel size still $^{1}/_{3}$ of the new aperture size, to replicate what would be seen when observing at that resolution. Therefore, the aperture diameter at each scale is the size of the effective beam FWHM at that scale. We spaced the apertures in the equivalent manner as shown in Figure~\ref{AppendixFig:IRAS_Sim_Apertures}, such that the sky regions to each side of the target position each sampled 3 FWHM at that simulated scale.

\begin{figure}
\centering
\includegraphics[width=0.475\textwidth]{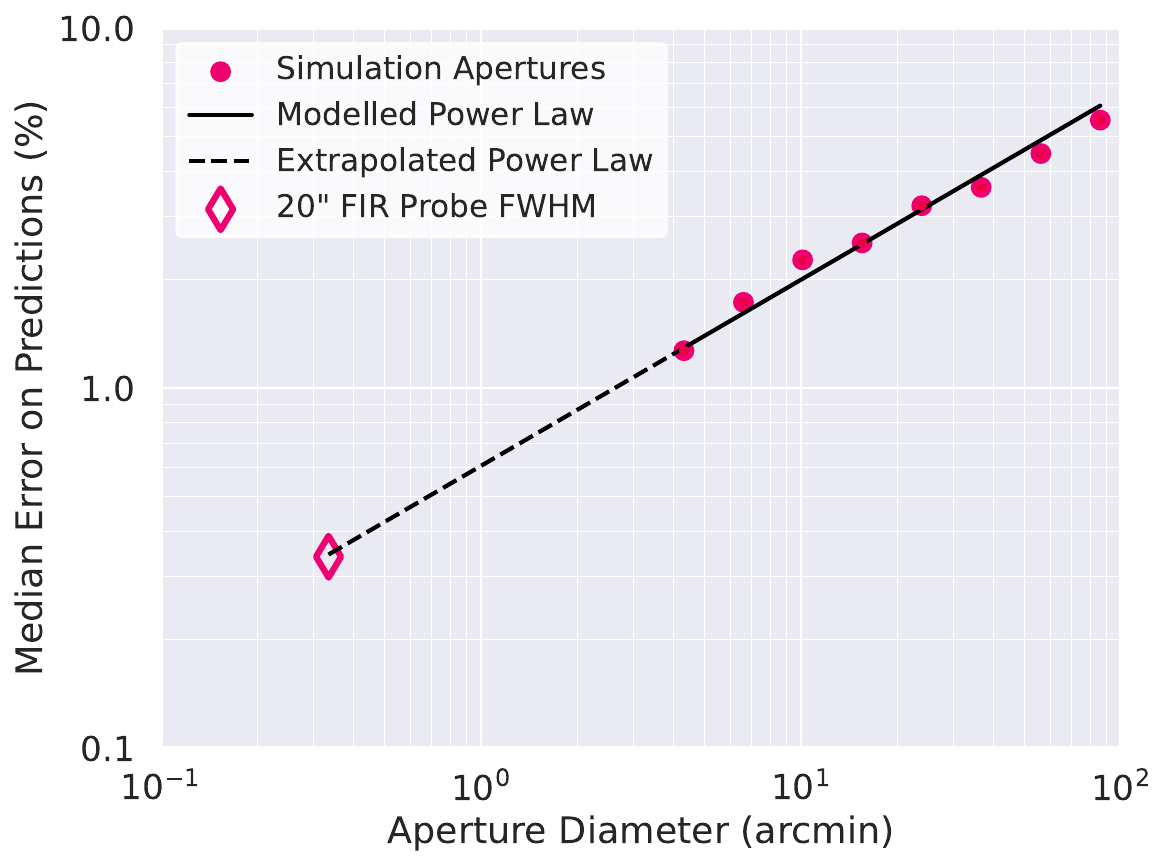}
\caption{Plot of the relationship between size of sky aperture, and median error on predicted value of sky brightness. Solid points show median error when using a range of aperture sizes for our IRAS simulations. Best fit power law to these points is plotted with the solid line; the dotted line extrapolates this to smaller aperture sizes. The predicted error on sky brightness predictions for 20\arcsec\ apertures is shown with the hollow diamond point.}
\label{AppendixFig:IRAS_Sim_Trend}
\end{figure}

For each aperture size, we repeated the entire simulation process detailed above, to find the median error in estimates on the diffuse Milky Way emission at that scale. The median error on the predicted sky emission for the simulations conducted at each aperture are shown in Figure~\ref{AppendixFig:IRAS_Sim_Trend}. It is clear that there is a very strong trend in how the error on the emission brightness predictions varies with the size of the aperture. We used 50 Monte Carlo bootstrap resampling (with replacement) to estimate the uncertainty on the median error for each aperture size; the uncertainties so small that the error bars are smaller than the datapoints as plotted in Figure~\ref{AppendixFig:IRAS_Sim_Trend}.

We found the least-squares best-fit power law to the relationship between the median error measured at each aperture scale, which has a power law index of $0.502 \pm 0.009$, with an intercept of $0.601\% \pm 0.021\%$ at 1\arcmin. Uncertainties on the power law fit were determined via 100 Monte Carlo bootstrap fits, in which each datapoint was permuted according to its uncertainty, following a Gaussian distribution. The best-fit power law is included in Figure~\ref{AppendixFig:IRAS_Sim_Trend}.

Our modelled power law allows us to extrapolate what typical error we would {\it expect}, when applying this emission-subtraction method at the 20\arcsec\ (or better) angular resolution of the proposed future FIR facilities. Specifically, we find that the median error on the estimation and subtraction of Milky Way emission at 20\arcsec\ scales will be {\bf only 0.33\%} $\pm$ 0.02\%. This is plotted in Figure~\ref{AppendixFig:IRAS_Sim_Trend}. \textcolor{black}{It has been established from \hersc\ and other data that the Galactic ISM exhibits fractal, scale-invariant geometry over a very wide range of angular scales \citep{Ossenkopf-Okada2019A,Roy2019A,Robitaille2019F}. So we can have a reasonable expectation that extrapolating to angular scale below that of IRAS, down to the \textless10\arcsec\ scales of interest, is reasonable .}

\needspace{3\baselineskip} \section{Catalog of Candidate Sightlines} \label{AppendixSection:Catalog}

Here we present our catalog of candidate sightlines. The full catalog tables are provided online as supplementary materials in machine-readable format. Here we give descriptions of the catalog tables, with example rows provided for explanatory purposes.

 In Table~\ref{AppendixTable:Catalog}, we provide the specific measured and predicted properties for each sightline. In Table~\ref{AppendixTable:Future_Facilities}, we provide the estimated observational parameters for each sightline for the proposed future FIR facilities PRIMA, FIRSST, and {\it Origins}, as described in Section~\ref{Section:Prospects}. 

\newpage
\begin{longtable*}{lrrrrrrrrrrr}
\caption{Catalog of our candidate sightlines for observing Milky Way \CII\ absorption. Each sightline is towards a FIR-bright nearby galaxy. The coordinates are of the FIR-brightest point used for the sightline, as per Section~\ref{Subsubsection:Background_Sources}; this is not necessarily the exact cener of the background galaxy. $F_{\it cont}$ is the 160\,\micron\ flux density of the background source, measured within the defined aperture. $\Delta F_{\it cont}$ is the reduction in flux density expected at the wavelength absorption is greatest. \CII $\frac{\rm emit}{\rm absorb}$ is the predicted ratio between Milky Way \CII\ emission and absorption for the sightline. Sub Err is the median error on the sky subtraction, estimated from our IRAS simulations, using the Gaussian Process Regression method. All other terms are defined in the text. The values of $N_{\rm C^+}$, $\int\tau_{\rm [CII]}dv$, $\tau_{\rm [CII]}^{\it (peak)}$, $\Delta F_{\it cont}$, and \CII $\frac{\rm emit}{\rm absorb}$, are all predictions. Here we provide 10 example rows to indicate the format and content; the table in its entirety is published online in machine-readable format: \url{https://iopscience.iop.org/article/10.3847/1538-3881/add40f\#ajadd40ft3}.}\label{AppendixTable:Catalog}\\
\toprule\toprule
\multicolumn{1}{c}{Name} &
\multicolumn{1}{c}{RA} &
\multicolumn{1}{c}{Dec} &
\multicolumn{1}{c}{$N_{\rm H}$} &
\multicolumn{1}{c}{$f_{\it mol}$} &
\multicolumn{1}{c}{$N_{\rm C^+}$} &
\multicolumn{1}{c}{$\int\tau_{\rm [CII]}dv$} &
\multicolumn{1}{c}{$\tau_{\rm [CII]}^{\it (peak)}$} & 
\multicolumn{1}{c}{$F_{\it cont}$} &
\multicolumn{1}{c}{$\Delta F_{\it cont}$} & 
\multicolumn{1}{c}{\CII$\frac{\rm emit}{\rm absorb}$} &
\multicolumn{1}{c}{Sub Err}\\
\multicolumn{1}{c}{} &
\multicolumn{1}{c}{(deg)} &
\multicolumn{1}{c}{(deg)} &
\multicolumn{1}{c}{(${\rm cm^{-2}}$)} &
\multicolumn{1}{c}{} &
\multicolumn{1}{c}{(${\rm cm^{-2}}$)} &
\multicolumn{1}{c}{(${\rm km\,s^{-1}}$)} &
\multicolumn{1}{c}{} & 
\multicolumn{1}{c}{(Jy)} &
\multicolumn{1}{c}{(Jy)} & 
\multicolumn{1}{c}{} &
\multicolumn{1}{c}{(\%)}\\
\cmidrule(lr){1-12}
NGC 0150 & 8.560 & -27.808 & 1.83$\times10^{20}$ & 0.001 & 3.76$\times10^{16}$ & 0.269 & 0.015 & 3.11 & -0.048 & 40.4 & 1.8\\
NGC 0253 & 11.884 & -25.292 & 3.71$\times10^{20}$ & 0.140 & 7.47$\times10^{16}$ & 0.533 & 0.012 & 310.13 & -3.698 & 0.1 & 1.4\\
NGC 0274 & 12.765 & -7.066 & 4.63$\times10^{20}$ & 0.001 & 9.25$\times10^{16}$ & 0.660 & 0.075 & 1.69 & -0.122 & 47.8 & 1.9\\
NGC 0289 & 13.175 & -31.207 & 1.89$\times10^{20}$ & 0.057 & 3.88$\times10^{16}$ & 0.277 & 0.022 & 2.62 & -0.057 & 46.6 & 0.9\\
NGC 0300 & 13.767 & -37.629 & 9.97$\times10^{20}$ & 0.000 & 1.95$\times10^{17}$ & 1.391 & 0.017 & 0.46 & -0.008 & 33.2 & 2.5\\
NGC 0337 & 14.960 & -7.582 & 7.99$\times10^{20}$ & 0.075 & 1.57$\times10^{17}$ & 1.122 & 0.126 & 1.75 & -0.207 & 50.9 & 2.6\\
NGC 0450 & 18.887 & -0.862 & 3.60$\times10^{20}$ & 0.000 & 7.24$\times10^{16}$ & 0.517 & 0.047 & 0.41 & -0.019 & 166.1 & 2.3\\
NGC 0520 & 21.146 & 3.792 & 3.29$\times10^{20}$ & 0.007 & 6.65$\times10^{16}$ & 0.475 & 0.034 & 12.97 & -0.439 & 9.5 & 1.3\\
NGC 0586 & 22.865 & -6.932 & 3.64$\times10^{20}$ & 0.040 & 7.32$\times10^{16}$ & 0.523 & 0.075 & 0.33 & -0.024 & 205.7 & 1.5\\
NGC 0613 & 23.575 & -29.420 & 1.63$\times10^{20}$ & 0.045 & 3.36$\times10^{16}$ & 0.240 & 0.024 & 8.08 & -0.192 & 16.1 & 2.1\\
\bottomrule
\end{longtable*}

\newpage
\begin{longtable*}{lrrrrrrrr}
\caption{Predicted observation parameters for our candidate sightlines for observing Milky Way \CII\ absorption, for the proposed future FIR facilities FIRSST, PRIMA, SALTUS, and {\it Origins}. For each facility, we give the predicted opacity of the \CII\ absorption feature, within the velocity resolution element where absorption peaks, and the predicted integration time necessary to detect the \CII\ absorption feature at S/N=10. Here we provide 10 example rows to indicate the format and content; the table in its entirety is published online in machine-readable format: \url{https://iopscience.iop.org/article/10.3847/1538-3881/add40f\#ajadd40ft4}.}\label{AppendixTable:Future_Facilities}\\
\toprule\toprule
\multicolumn{1}{c}{Name} &
\multicolumn{2}{c}{FIRSST} &
\multicolumn{2}{c}{PRIMA} &
\multicolumn{2}{c}{SALTUS} &
\multicolumn{2}{c}{\it Origins} \\
\cmidrule(lr){2-3}
\cmidrule(lr){4-5}
\cmidrule(lr){6-7}
\cmidrule(lr){8-9}
\multicolumn{1}{c}{} &
\multicolumn{1}{c}{$\tau_{\rm [CII]}^{\it (peak)}$} &
\multicolumn{1}{c}{$t_{\it 10\sigma}$ (sec)} &
\multicolumn{1}{c}{$\tau_{\rm [CII]}^{\it (peak)}$} &
\multicolumn{1}{c}{$t_{\it 10\sigma}$ (sec)} &
\multicolumn{1}{c}{$\tau_{\rm [CII]}^{\it (peak)}$} &
\multicolumn{1}{c}{$t_{\it 10\sigma}$ (sec)} &
\multicolumn{1}{c}{$\tau_{\rm [CII]}^{\it (peak)}$} &
\multicolumn{1}{c}{$t_{\it 10\sigma}$ (sec)} \\
\cmidrule(lr){1-9}
NGC 0150 & 0.015 & 4.89$\times10^{1}$ & 0.003 & 5.78$\times10^{1}$ & 0.015 & 3.61$\times10^{0}$ & 0.009 & 2.32$\times10^{-2}$\\
NGC 0253 & 0.012 & 1.24$\times10^{-3}$ & 0.005 & 1.47$\times10^{-3}$ & 0.012 & 1.29$\times10^{-4}$  & 0.011 & 5.92$\times10^{-7}$\\
NGC 0274 & 0.070 & 2.85$\times10^{1}$ & 0.000 & 3.38$\times10^{1}$ & 0.075 & 2.11$\times10^{0}$ & 0.015 & 1.34$\times10^{-2}$\\
NGC 0289 & 0.022 & 6.50$\times10^{1}$ & 0.000 & 7.67$\times10^{1}$ & 0.022 & 4.80$\times10^{0}$ & 0.014 & 3.07$\times10^{-2}$\\
NGC 0300 & 0.017 & 8.31$\times10^{1}$ & 0.003 & 9.81$\times10^{1}$ & 0.017 & 2.80$\times10^{1}$ & 0.017 & 3.93$\times10^{-2}$\\
NGC 0337 & 0.120 & 9.57$\times10^{0}$ & 0.000 & 1.13$\times10^{1}$ & 0.126 & 7.06$\times10^{-0}$ & 0.065 & 4.50$\times10^{-3}$\\
NGC 0450 & 0.046 & 7.82$\times10^{2}$ & 0.008 & 9.23$\times10^{2}$ & 0.046 & 5.77$\times10^{1}$ & 0.037 & 3.70$\times10^{-1}$\\
NGC 0520 & 0.033 & 9.10$\times10^{-1}$ & 0.005 & 1.08$\times10^{0}$ & 0.034 & 6.74$\times10^{-2}$ & 0.012 & 4.30$\times10^{-4}$\\
NGC 0586 & 0.047 & 1.22$\times10^{3}$ & 0.007 & 1.45$\times10^{3}$ & 0.075 & 9.06$\times10^{1}$ & 0.024 & 5.77$\times10^{-1}$\\
NGC 0613 & 0.019 & 9.11$\times10^{1}$ & 0.001 & 1.08$\times10^{1}$ & 0.024 & 6.73$\times10^{-1}$ & 0.011 & 4.30$\times10^{-3}$\\
\bottomrule
\end{longtable*}


\bibliography{ChrisBib}{}
\bibliographystyle{aasjournal}



\end{document}